# The echo-enabled harmonic generation options for FLASH II

Haixiao Deng [1, 2, *], Winfried Decking [2], Bart Faatz [2]

*1. Shanghai Institute of Applied Physics, Jialuo Road 2019, 201800 Shanghai, China*
*2. Deutsches Elektronen-Synchrotron (DESY), Notkestrasse 85, 22603 Hamburg, Germany*

FLASH II is an upgrade to the existing free electron laser (FEL) FLASH. The echo-enabled harmonic generation (EEHG) scheme is proposed to be a potential seeding option of FLASH II. In this paper, the possibility of EEHG operation of FLASH II is investigated for the first time. With a combination of existing numerical codes, i.e. a laser-beam interaction code in an undulator (LBICU), a beam tracking code in a chicane (ELEGANT) and an universal FEL simulating code (GENESIS), the effects of beam energy chirp and coherent synchrotron radiation (CSR) on EEHG operation are studied as well. In addition, several interesting issues concerning EEHG simulation are discussed.

*E-mail address:* denghaixiao@sinap.ac.cn

# I.  INTROUDCTION

On the way to fully coherent, short-wavelength free electron laser (FEL), the recently proposed double-modulator FEL scheme, so-called echo-enabled harmonic generation (EEHG) [1~2], holds promising prospects for efficiently generate high harmonic density modulation with a relatively small energy spread. The schematic of EEHG and the typical electron beam phase space in EEHG process are shown in Figure 1. Two proof-of-principle experiments of the EEHG technique have been carried out at SDUV-FEL [3~4] and NLCTA [5] independently.

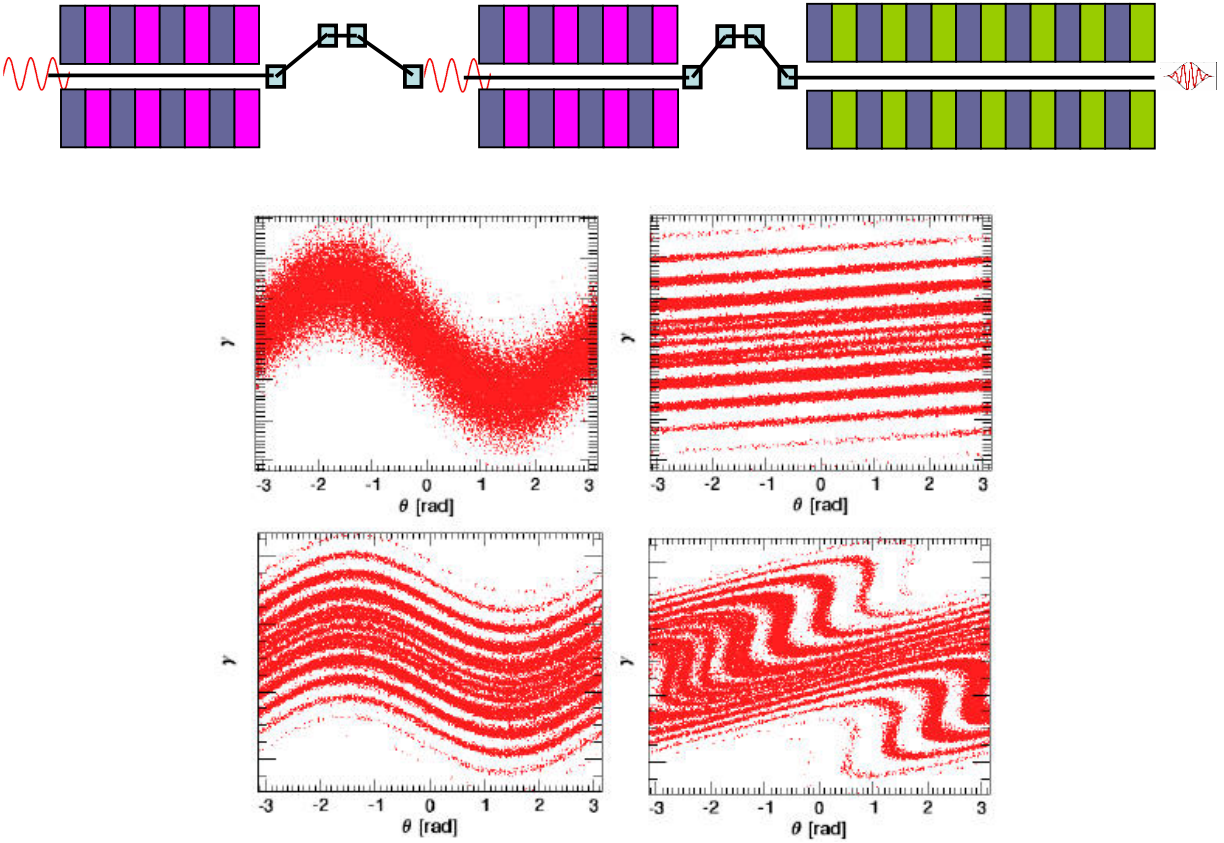

**Figure 1** Schematic EEHG layout and typical longitudinal phase space of the EEHG mechanism.

FLASH II is an upgrade of FLASH [6], where the electron beam is planed to be extracted to a new undulator line after the SRF linac. In addition to the self-amplified spontaneous emission (SASE) [7~12] operation as used for FLASH, FLASH II is considered to employ a seeded FEL scheme as well. The baseline design of FLASH II is two-stage cascaded high-gain harmonic generation (HGHG) [13~17] and one-stage high harmonic generation (HHG) amplification [18~21]. However, because of the remarkable performance promised by the EEHG theory, the EEHG scheme is considered to be an additional option for FLASH II.

In this paper, the possibility of EEHG operation of FLASH II is investigated. The effects of electron beam energy chirp and coherent synchrotron radiation (CSR) are studied with a combined three-dimensional numerical simulation.

## II.     EEHG OPTION FOR FLASH II

After installing a new superconductive RF module, the electron beam energy of the FLASH linac reaches 1.25GeV, opening the view through the water-window [22]. In this paper, three different electron beam energies, i.e. 0.7GeV, 1.0GeV and 1.22GeV are considered for FLASH II. If one considers a main undulator similar to FLASH, it will radiate at 13.1nm, 6.55nm and 4.37nm, respectively. It is corresponding to the $20^{th}$, $40^{th}$ and $60^{th}$ harmonics of a 262nm UV seed laser, which is usually generated from the $3^{rd}$ harmonic of a ~800nm Ti-sapphire laser.

One possible layout of the EEHG option for FLASH II is illustrated in Figure 2. Matching section A (MA) is used to optimize the initial TWISS function, especially small beta function, to minimum the emittance growth in the large chicane B1, and matching section B (MB) is used to match the FODO periodic beam envelope in the main undulator. The advantage of such an arrangement of MB is that, once the electron beam size is enlarged by the natural diffraction in the large chicane B1, it can be refocused in the $2^{nd}$ modulator M2. In more detail, MB usually should be more than a regular FODO lattice, as it is defined by the final radiator. Moreover, to ensure the dispersion closure of the large chicane B1, other small chicanes can be placed in two half-cell of the matching FODO lattice. However, it is worth stressing that the technical designs, including the final lattice structure and the seed laser coupling method, are out of the scope of this paper.

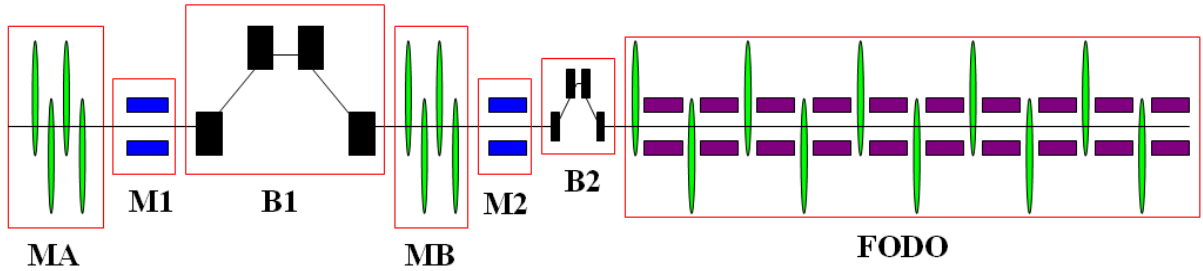

**Figure 2** Possible layout of EEHG option for FLASH II.

The main beam parameters of FLASH II are listed in Table 1.

**Table 1** The main beam parameters of FLASH II

| Beam energy | 0.7 / 1.0 / 1.22 GeV |
|---|---|
| Normalized emittance | 1.5 mm-mrad |
| Slice energy spread | 0.2 MeV |
| Peak current | 1.25 / 2.5 kA |
| Bunch charge | 1 nC |
| Beam size in Modulator | 60 um |

The main parameters of the large chicane B1 and the small chicane B2 are listed in Table 2.

**Table 2** The main parameters of the 1st chicane in EEHG option for FLASH II

|  | Chicane B1 | Chicane B2 |
|---|---|---|
| Dipole length | 0.4 m | 0.1m |
| Dipole field | 0 ~ 0.5 T | 0~0.2T |
| Drift between the 1st/2nd and 3rd/4th dipole | 2.0 m | 0.2m |
| Drift between the 2nd/3rd dipole | 0.4 m | 0.2m |

To obtain a significant density modulation in the higher harmonics, we choose the seed laser parameters and make the maximum beam energy modulation to be 3 times of the slice energy spread in both M1 and M2. Moreover, the seed laser radius is much larger than the electron beam radius in both M1 and M2. In the simulation, the 262nm seed laser radius is 0.6mm, while the peak power is in about 1.2~1.5GW for the 3 different cases.

A numerical solution is developed to calculate the EEHG progress. For details, please go to the Appendixes. In order to obtain the working point of EEHG setup, an ideal Gaussian electron beam is considered in this section, and the CSR effects in chicanes are not taken into accounts. Since the seed laser induced energy spread was determined, in the following we only scan the dispersive strength of the two chicanes to optimize the EEHG setup and find its working point.

### a) 20th harmonic case

Figure 3 shows the 20th harmonic bunching factor of the electron beam as the chicane parameters vary. From Figure 3, the bunching factor at the 20th harmonic of the seed laser (i.e. 13.1nm) is optimized at a $R_{56}^{(1)}$ of 1.08mm, corresponding an 0.884 degree bending angle of the 1st chicane.

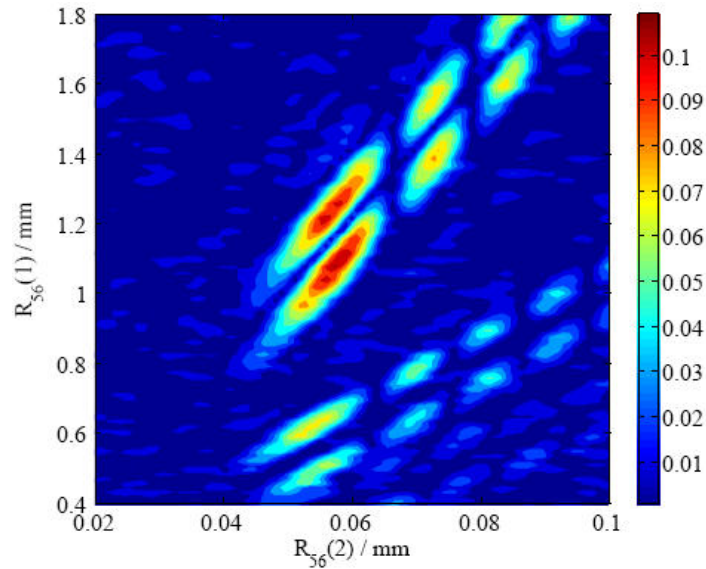

**Figure 3** The 20th harmonic bunching .vs. the dispersion of the 1st chicane and the 2nd chicane in EEHG setup of FLASH II.

The two lower parallel areas correspond to an electron beam modulation on the sub-2 harmonic of the 13.1nm, i.e. 26.2nm. Figure 4 shows the 20$^{th}$ harmonic bunching variation with the 2$^{nd}$ chicane parameter when the $R_{56}^{(1)}$ of 1$^{st}$ chicane is set to be 1.08mm, and it suggests an optimized $R_{56}^{(2)}$ of 0.0576 mm. Figure 3 and Figure 4 are consistent with the well-known EEHG theory.

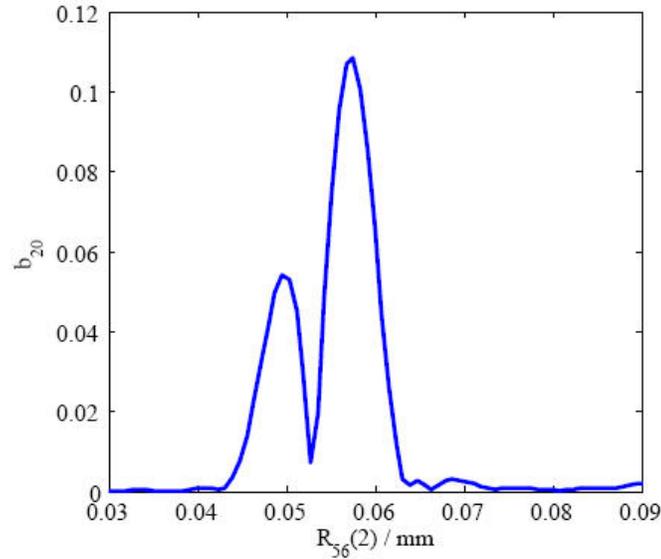

**Figure 4** The 20$^{th}$ harmonic bunching vs. the longitudinal dispersion of the 2$^{nd}$ chicane in the EEHG setup, when the bending angle of the 1$^{st}$ chicane is optimized.

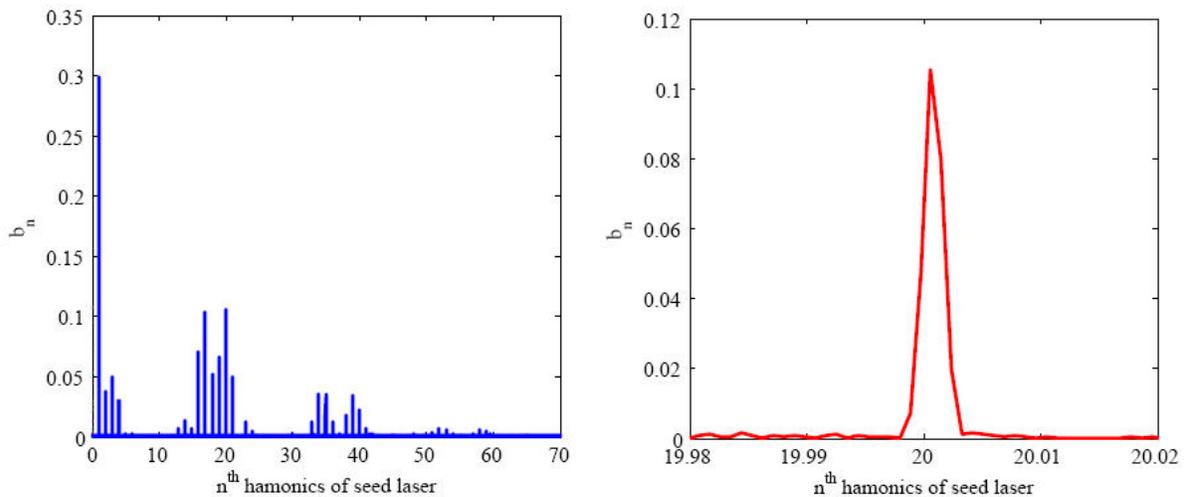

**Figure 5** The current spectrum of the modulated electron beam when EEHG setup is optimized at the 20$^{th}$ harmonic of the seed laser.

Figure 5 gives the beam current spectrum, when modulated by the optimized EEHG setup. Here a Gaussian beam with 1nC bunch charge, 2.5kA peak current was used.

### b) 40$^{th}$ harmonic case

Figure 6 shows the 40$^{th}$ harmonic bunching factor of the electron beam as the chicane parameters vary. From Figure 6, the bunching factor at the 40$^{th}$ harmonic of the seed laser (i.e. 6.55nm) is optimized at a $R_{56}^{(1)}$ of 2.86mm, corresponding an 1.44 degree bending angle of the 1$^{st}$ chicane.

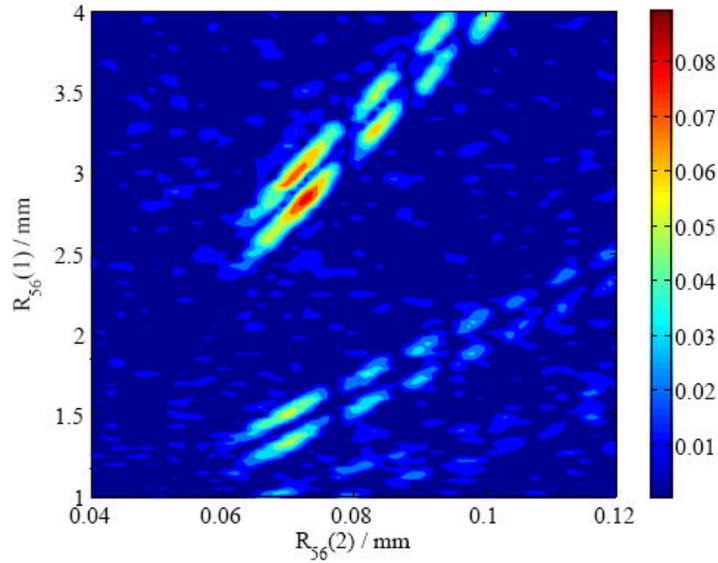

**Figure 6** The 40$^{th}$ harmonic bunching .vs. the dispersion of the 1$^{st}$ chicane and the 2$^{nd}$ chicane in EEHG setup of FLASH II.

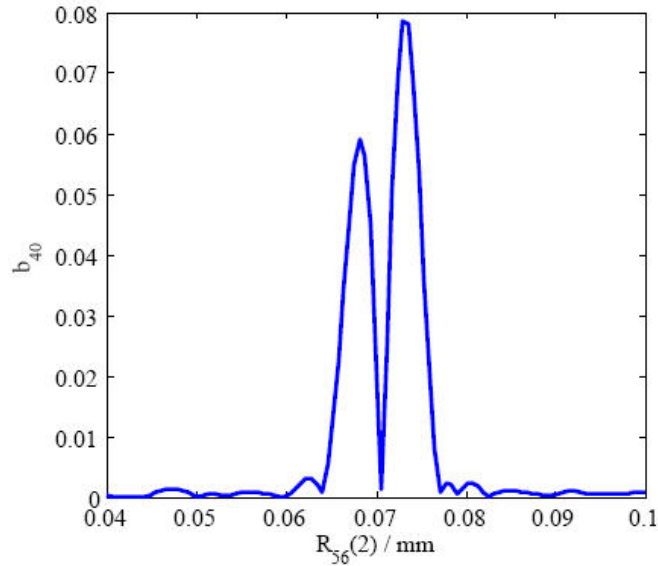

**Figure 7** The 40$^{th}$ harmonic bunching vs. the dispersion of the 2$^{nd}$ chicane in the EEHG setup, when the bending angle of the 1$^{st}$ chicane is optimized.

The two lower parallel areas correspond to the electron beam being effectively modulated on the sub-2 harmonic of the 6.55nm, i.e. 13.1nm. Figure 7 shows the 40$^{th}$ harmonic bunching variation with the 2$^{nd}$ chicane parameter when the 1$^{st}$ chicane is optimized, which suggests an optimized $R_{56}^{(2)}$ of 0.0730mm. Figure 6 and Figure 7 are consistent with the well-known EEHG theory. Figure 8 gives the beam current spectrum, when modulated by the optimized EEHG setup. Here a Gaussian beam with 1nC bunch charge, 2.5kA peak current was used.

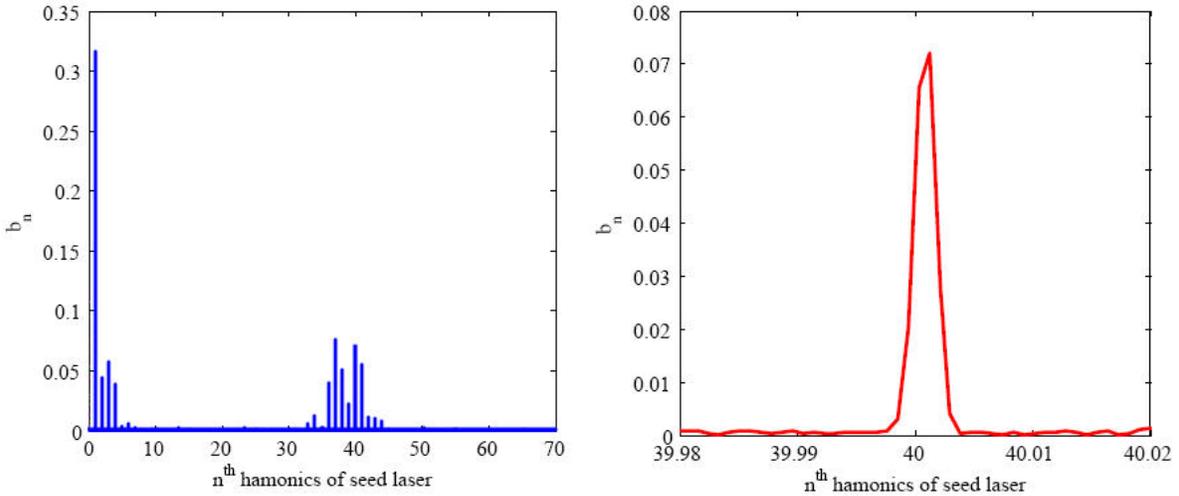

**Figure 8** The current spectrum of the modulated electron beam when the EEHG setup is optimized at the 40$^{th}$ harmonic of the seed laser.

c) **60$^{th}$ harmonic case**

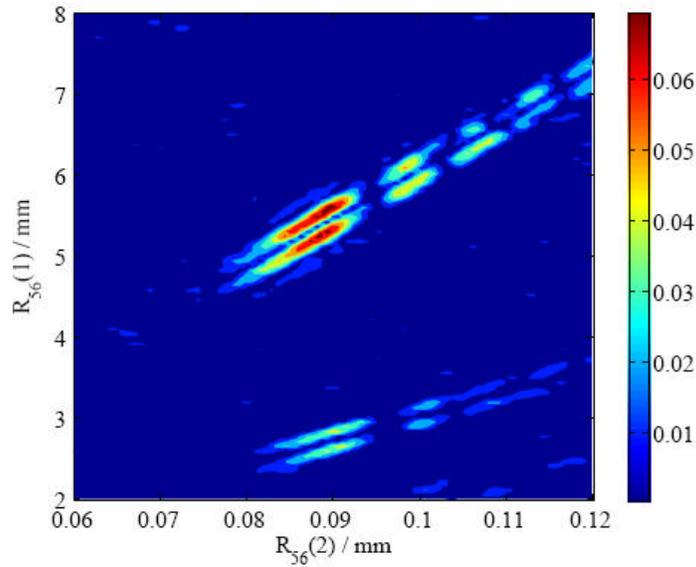

**Figure 9** The 60$^{th}$ harmonic bunching .vs. the dispersion of the 1$^{st}$ chicane and the 2$^{nd}$ chicane in EEHG setup of FLASH II.

Figure 9 shows the 60th harmonic bunching factor of the electron beam as the chicane parameters varies. From Figure 9, the bunching factor at the 60th harmonic of the seed laser (i.e. 4.37nm) is optimized at a $R_{56}^{(1)}$ of 5.25mm, corresponding an 1.95 degree bending angle of the 1st chicane.

The two lower parallel areas show where the electron beam is effectively modulated on the sub-2 harmonic of the 4.37nm, i.e. 8.74nm. Figure 10 shows the 60th harmonic bunching variation with the 2nd chicane parameter when the 1st chicane is optimized, which suggests an optimized $R_{56}^{(2)}$ of 0.0888mm. Figure 9 and Figure 10 are consistent with the well-known EEHG theory

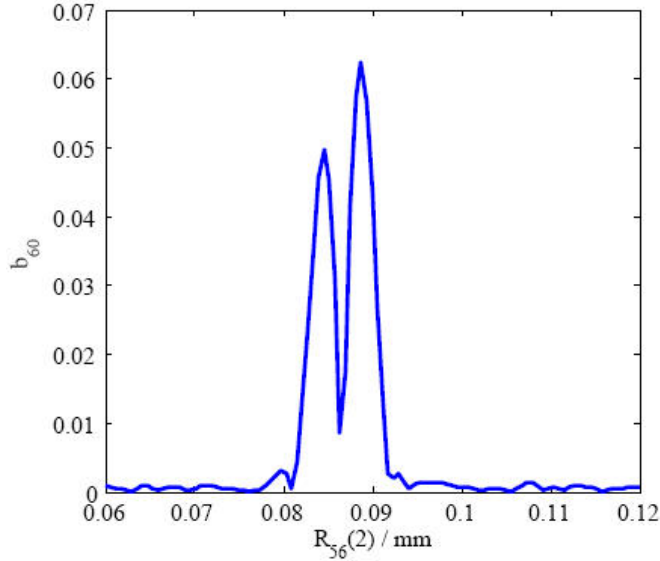

**Figure 10** The 60th harmonic bunching vs. the dispersion of the 2nd chicane in EEHG setup, when the bending angle of the 1st chicane is optimized.

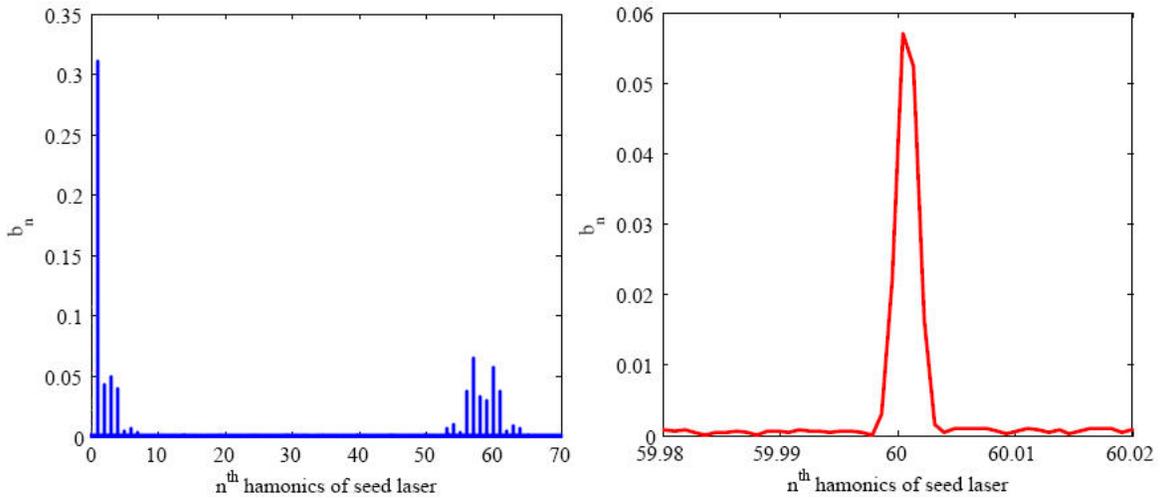

**Figure 11** The current spectrum of the modulated electron beam when the EEHG setup is optimized at the 60th harmonic of the seed laser.

Figure 11 gives the beam current spectrum when modulated by the optimized EEHG setup. Here a Gaussian beam with 1nC bunch charge, 2.5kA peak current was used.

In addition, a 1.25kA peak current case has also been simulated for the 3 different harmonics. Since the CSR effects are not taken into accounts in this section, the results are similar to the presented 2.5kA peak current case.

### III. ENERGY CHIRP EFFECTS

For the sake of high-ratio longitudinal compression, a large beam energy chirp is introduced in FLASH. The energy chirp cannot be removed in the limited 1.3GHz superconductive RF linac. Figure 12 shows the longitudinal phase space of the electron beam in FLASH linac, obtained from start2end simulations for 1nC bunch charge [23].

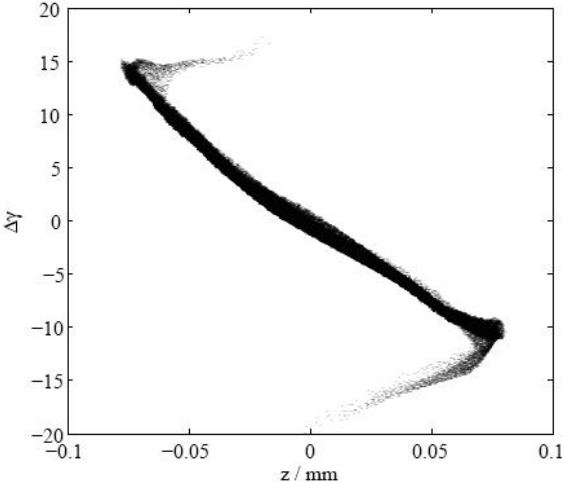

**Figure 12** Start2end simulation results of the FLASH beam at the exit of linac.

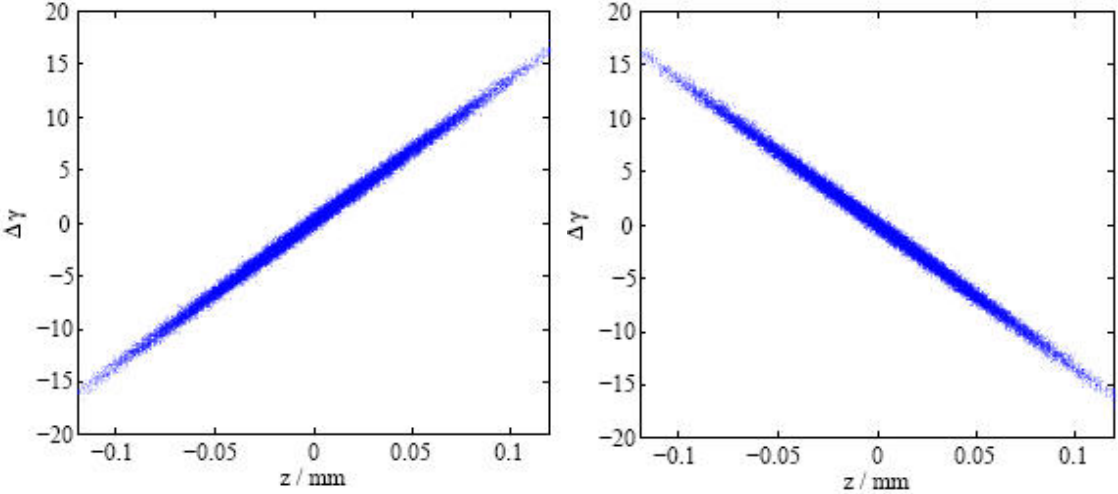

**Figure 13** Gaussian beam with a positive (left) and a negative (right) beam energy chirp.

In this section, a linear energy chirp is introduced to the ideal Gaussian beam to reproduce the start2end simulation results, and the sensitivity of EEHG on this chirp is studied. As shown in Figure 13, a positive and a negative beam energy chirp is investigated for the abovementioned three EEHG options of FLASH II.

### a) $20^{th}$ harmonic case

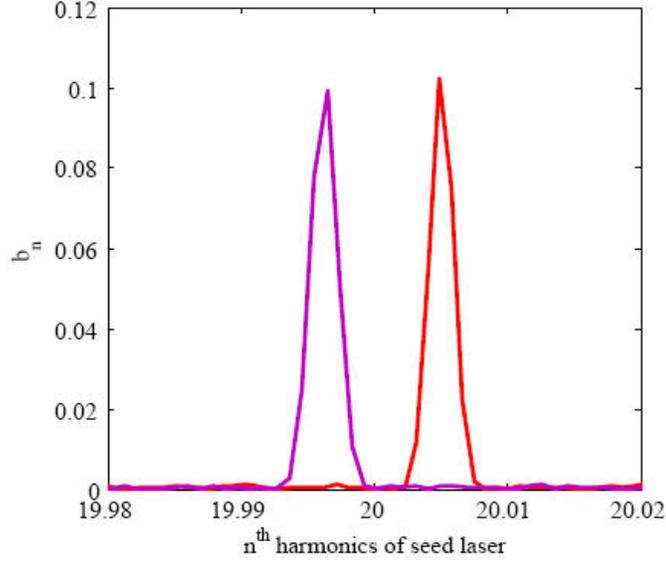

**Figure 14** EEHG wavelength shift induced by the beam energy chirp in the $20^{th}$ harmonic option. The purple represents the positive energy chirp and the red represents the negative energy chirp.

In absence of CSR effects, the simulation result for the $20^{th}$ harmonic EEHG option is illustrated in Figure 14. When the electron beam has a linearly correlated energy spread of ±10, the optimized EEHG wavelength varies about ±0.003nm from the 13.1nm. However, the amplitude of the bunching factor remains at a level of 0.1.

### b) $40^{th}$ harmonic case

In absence of CSR effects, the simulation result for the $40^{th}$ harmonic EEHG option is illustrated in Figure 15. When the electron beam has a linearly correlated energy spread of ±10, the optimized EEHG wavelength varies about ±0.001nm from the 6.55nm. However, the amplitude of the bunching factor remains at a level of 0.06.

### c) $60^{th}$ harmonic case

In absence of CSR effects, the simulation result for the $60^{th}$ harmonic EEHG option is illustrated in Figure 15. When the electron beam has a linearly correlated energy spread of ±10, the optimized EEHG wavelength varies less than ±0.001nm from the 4.37nm. However, the amplitude of the bunching factor remains at a level of 0.05.

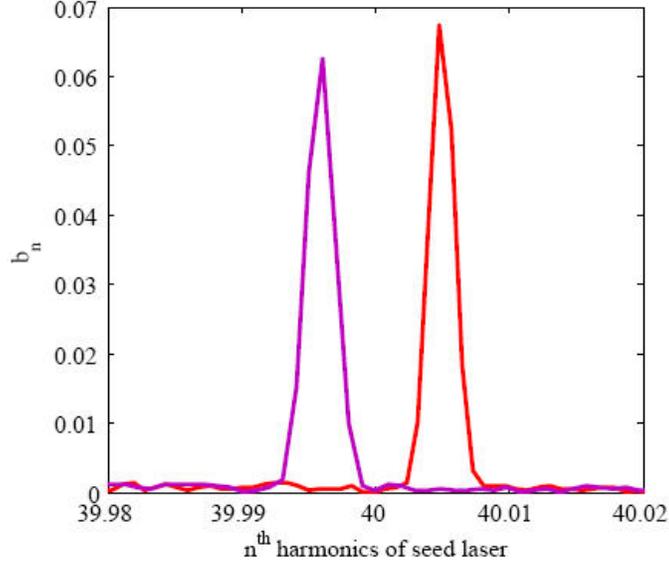

**Figure 15** EEHG wavelength shift induced by the beam energy chirp in the 40th harmonic option. The purple represents the positive energy chirp and the red represents the negative energy chirp.

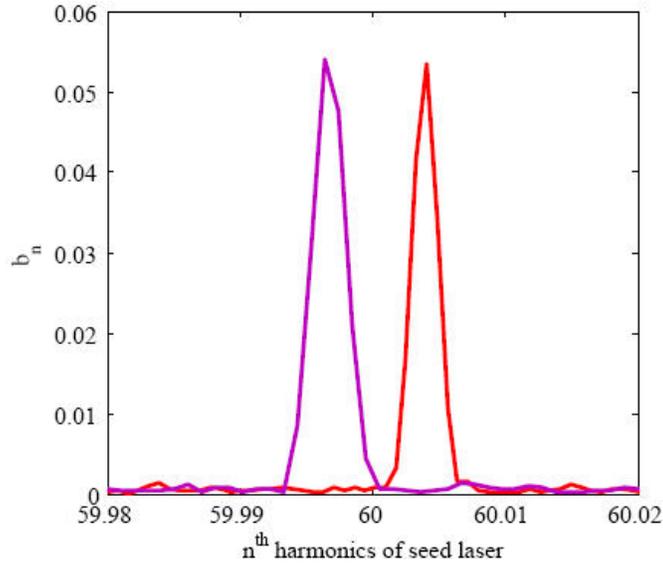

**Figure 16** EEHG wavelength shift induced by the beam energy chirp in the 60th harmonic option. The purple represents the positive energy chirp and the red represents the negative energy chirp.

The effects of a linear beam energy chirp on echo microbunching and EEHG FEL have been theoretically [24] and numerically studied in one dimension [2, 5 and 24]. A three-dimensional simulation of beam energy chirp effects is presented here, which is well consistent with the theory [24]. Despite the strong chicane, the microbunching wavelength shift is very small in the studied cases of the EEHG option for FLASH II.

Since we calculate the density modulation of the whole electron bunch, the expected theoretical bandwidth of the microbunching is about $10^{-5}$, which is consistent with the optimized microbunching bandwidth in simulation. Thus, a microbunching wavelength shift which is larger than the bandwidth can be observed. Moreover, it is worth stressing that a different energy chirp results in a different microbunching wavelength shift. This observation will be helpful to understand the CSR effects on the EEHG mechanism in the following discussions.

## IV. CSR- INDUCED DEGRADATION

The EEHG mechanism requires a strong chicane if used up to high harmonics, then strong CSR effects will definitely degrade the EEHG performances. Since the electron beam envelope is optimized for minimum emittance growth when passing through the first strong chicane in the EEHG setup by the matching section MA, the projected emittance growth is well-controlled. As shown in Figure 17, it is less than 10% even for the 2.5kA peak current case in the $60^{th}$ harmonic EEHG option for FLASH II.

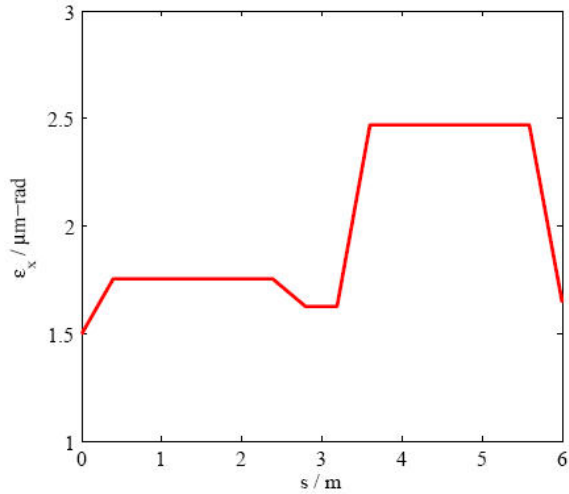

**Figure 17** The projected emittance growth in the first chicane of the EEHG setup, where the EEHG is optimized for the $60^{th}$ harmonic and the peak current is 2.5kA.

Thus, we mainly concentrate on the density modulation efficiency in this section.

### a) 20th harmonic case

The CSR affected microbunching spectra of the 20th harmonic EEHG option are plotted in Figure 18 and 19. In the 1.25kA peak current case, the maximum bunching factor around the 13.1nm wavelength reduces from 0.105 to 0.058, while it reduces to 0.042 in the 2.5kA peak current case. In both cases, the beam microbunching bandwidths are broadened by the CSR effects. However, in the higher peak current situation, stronger CSR force broadens the bandwidth more significantly.

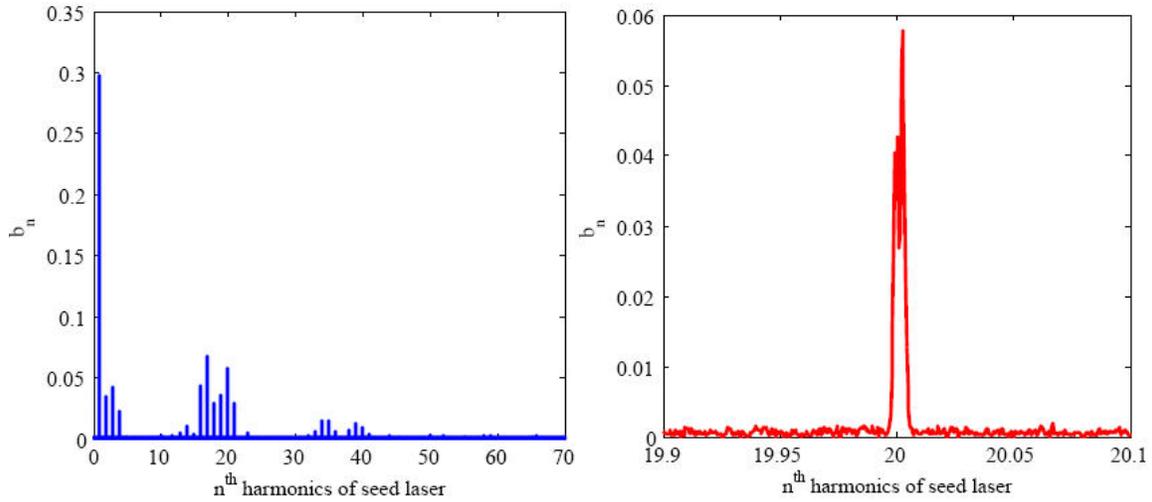

**Figure 18** EEHG microbunching spectra of the 20th harmonic option for FLASH II, where the CSR effects in the first chicane are taken into account. The beam peak current is 1.25kA.

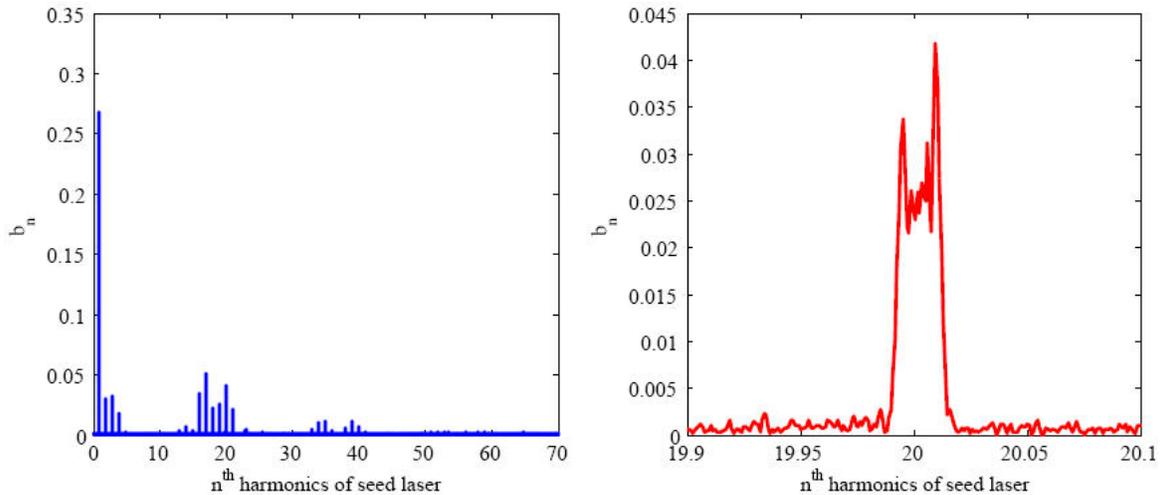

**Figure 19** EEHG microbunching spectra of the 20th harmonic option for FLASH II, where the CSR effects in the first chicane are taken into account. The beam peak current is 2.5kA.

### b) 40<sup>th</sup> harmonic case

The CSR affected microbunching spectra of the 40$^{th}$ harmonic EEHG option are plotted in Figure 20 and 21. In the 1.25kA peak current case, the maximum bunching factor around the 6.55nm wavelength reduces from 0.072 to 0.034, while it reduces to 0.022 in the 2.5kA peak current case. In both cases, the beam microbunching bandwidths are broadened by the CSR effects. However, in the higher peak current situation, stronger CSR force broadens the bandwidth more significantly.

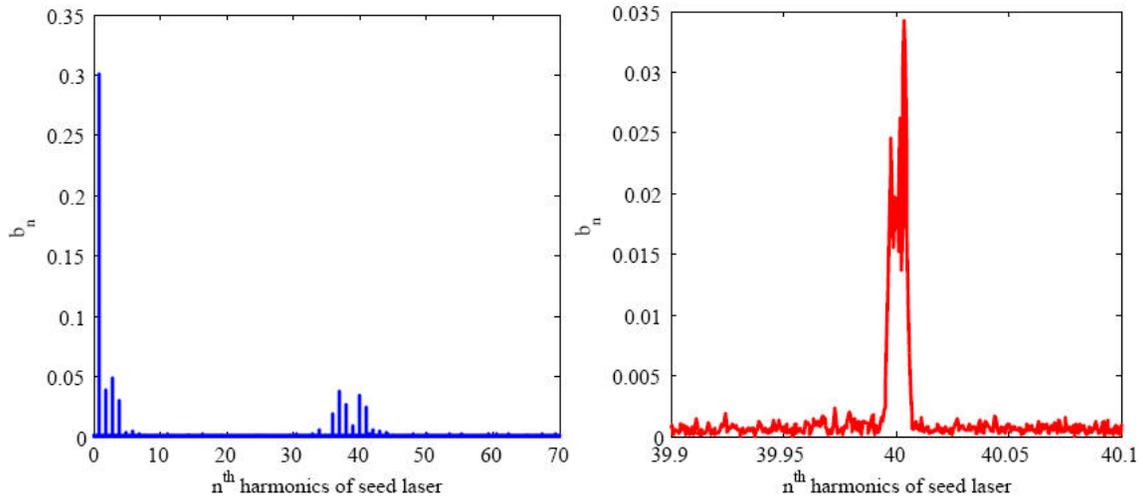

**Figure 20** EEHG microbunching spectra of the 40$^{th}$ harmonic option for FLASH II, where the CSR effects in the first chicane are taken into account. The beam peak current is 1.25kA.

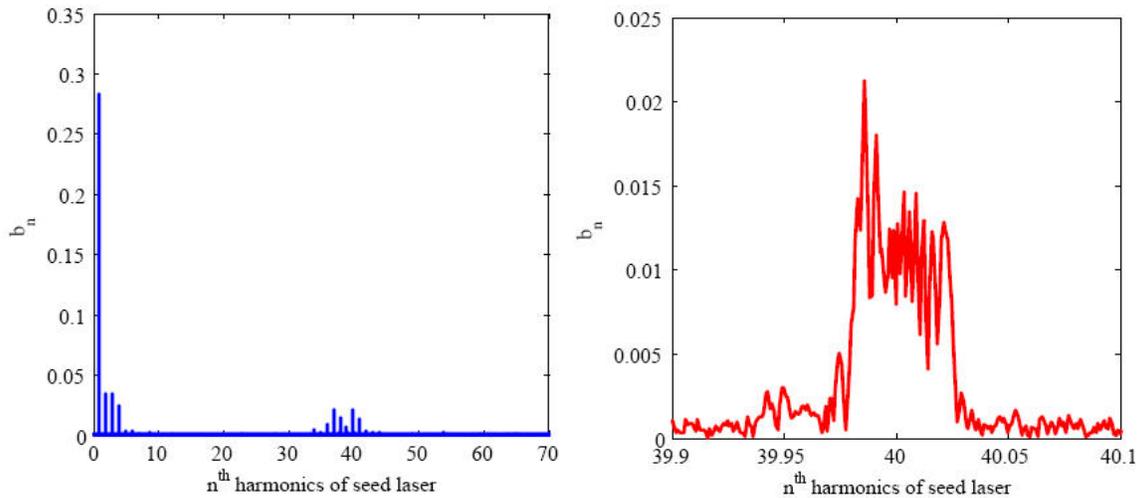

**Figure 21** EEHG microbunching spectra in the 40$^{th}$ harmonic option for FLASH II, where the CSR effects in the first chicane are taken into account. The beam peak current is 2.5kA.

### c) 60th harmonic case

The CSR affected microbunching spectra of the 60th harmonic EEHG option are plotted in Figure 22 and 23. In the 1.25kA peak current case, the maximum bunching factor around the 4.37nm wavelength reduces from 0.057 to 0.018, while it reduces to 0.012 in the 2.5kA peak current case. In both cases, the beam microbunching bandwidths are broadened by the CSR effects. However, in the higher peak current situation, stronger CSR force broadens the bandwidth more significantly.

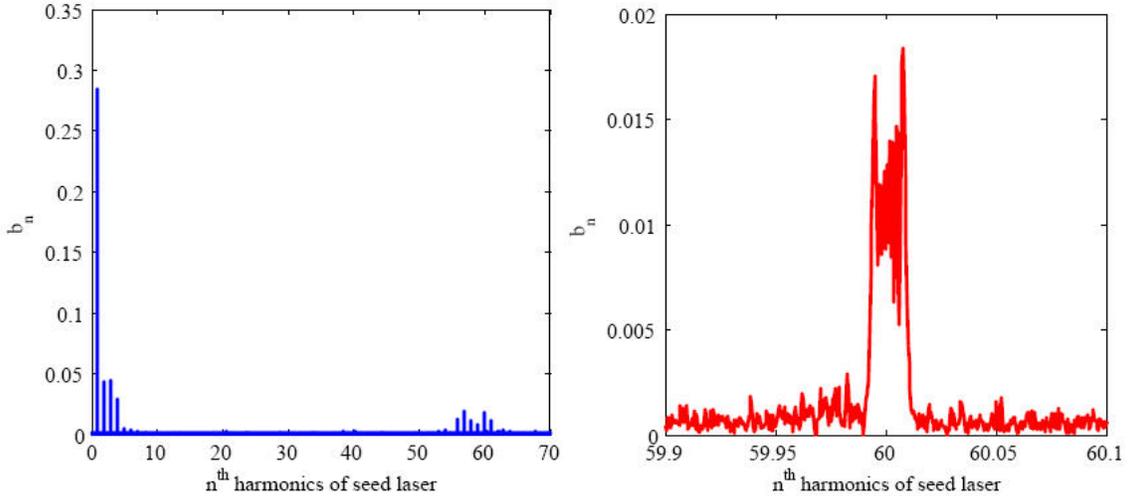

**Figure 22** EEHG microbunching spectra of the 60th harmonic option for FLASH II, where the CSR effects in the first chicane are taken into account. The beam peak current is 1.25kA.

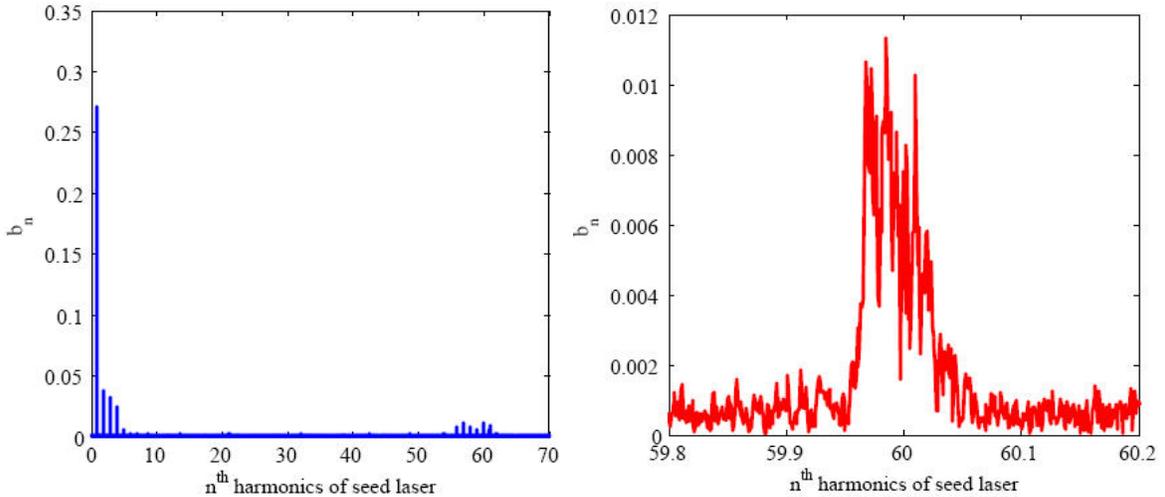

**Figure 23** EEHG microbunching spectra of the 60th harmonic option for FLASH II, where the CSR effects in the first chicane are taken into account. The beam peak current is 2.5kA.

The CSR affects the EEHG microbunching as follows. The CSR that is emitted in the strong dipole field of the chicane bending magnets interacts with the electron beam and yields an energy modulation of the electron beam. The CSR-induced energy modulation varies along the longitudinal axis of the beam, resulting in a variation of the energy chirp in different part of the beam, as seen in Figure 24. As described in section III and in other refs [24~25], a linear energy chirp results in a microbunching wavelength shift, and then different parts of the electron beam shift to different microbunching wavelength. Thus the microbunching bandwidth is broadened and the microbunching amplitude is degraded.

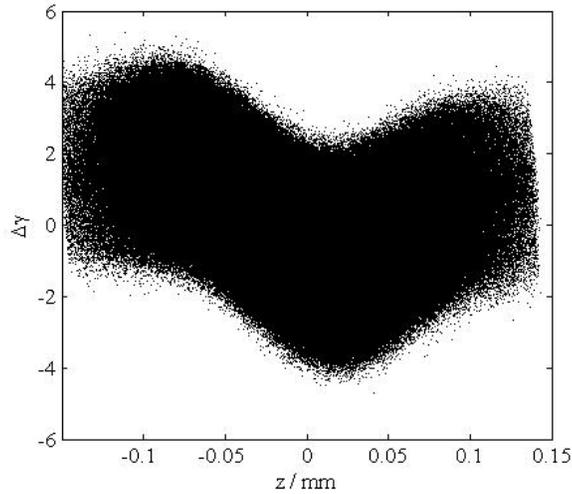

**Figure 24** The longitudinal electron beam phase space affected by the CSR field in the case of $60^{th}$ harmonic of EEHG setup for FLASH II.

According to the simulation, the microbunching performance of the whole electron beam, i.e. the projected bunching factor, is significantly degraded by CSR. However, if we consider only the typical FEL cooperation length instead of the whole bunch length, taking into account that the FEL radiation growth is only determined by the beam behavior within an FEL cooperation length, the initial microbunching is expected to be much larger within these short slices. In other words, the CSR effects degrade the effective microbunching bandwidth, but as long it is smaller than the FEL bandwidth, it shouldn't matter so much.

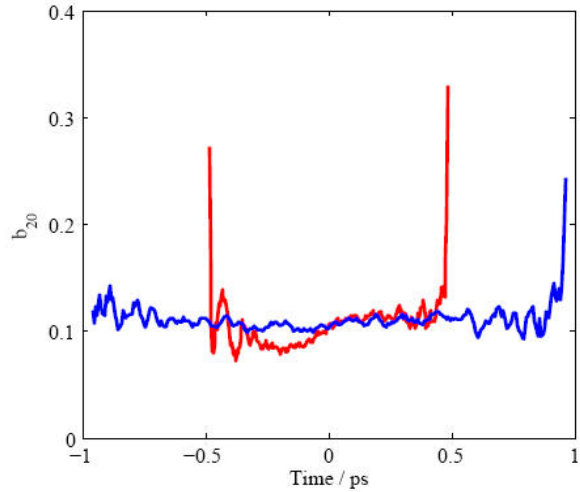

**Figure 25** Sliced 20[th] harmonic bunching factor at the entrance of the radiator, where the blue is the 1.25kA peak current case and the red is the 2.5kA peak current case.

Figure 25 shows the sliced bunching factor of the EEHG-modulated beam at the entrance of the radiator in the 20[th] harmonic EEHG setup. Although the bunching factor of the whole electron beam is reduced from 0.105 to 0.042 in the presence of the CSR effects, the local bunching factor of the electron beam is still at an order of 0.10, which will contribute to an initial coherent harmonic generation in the radiator.

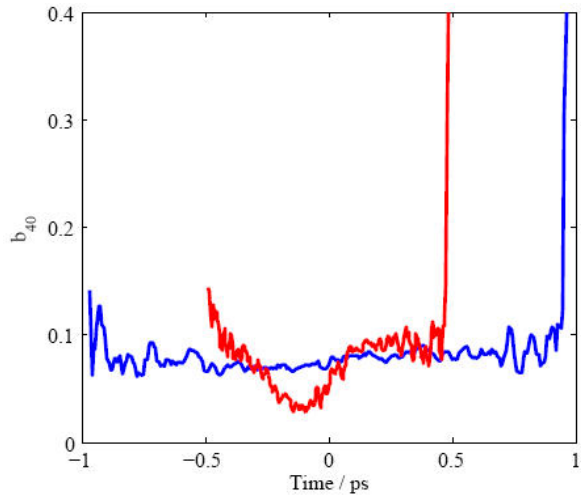

**Figure 26** Sliced 40[th] harmonic bunching factor at the entrance of the radiator, where the blue is the 1.25kA peak current case and the red is the 2.5kA peak current case.

Figure 26 shows the sliced bunching factor of the EEHG-modulated beam at the entrance of the radiator in the 40[th] harmonic EEHG setup. Although the bunching factor of the whole electron beam is reduced from 0.072 to 0.022 in the presence of the CSR effects, the local bunching factor of the electron beam is still over 0.05.

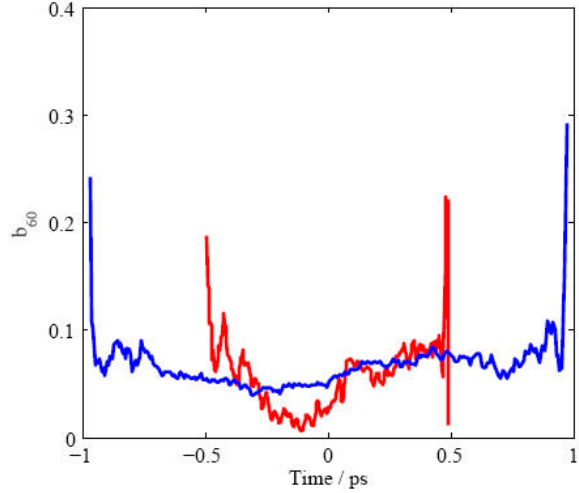

**Figure 27** Sliced 60<sup>th</sup> harmonic bunching factor at the entrance of the radiator, where the blue is the 1.25kA peak current case and the red is the 2.5kA peak current case.

Figure 27 shows the sliced bunching factor of the EEHG-modulated beam at the entrance of the radiator in the 60$^{th}$ harmonic EEHG setup. Although the bunching factor of the whole electron beam is reduced from 0.057 to 0.012 in the presence of the CSR effects, the local bunching factor of the electron beam is still strong enough in many beam slices.

## V. FEL PROPERTIES

In this section, we calculate the FEL performance of the EEHG-modulated beam in the radiator undulator of FLASH II. All the simulations include the CSR effects as described above. The same undulator as in FLASH is assumed. Table 3 lists the main parameters of EEHG option for FLASH II.

**Table 3** The main parameters of EEHG setup for FLASH II

|  | 20$^{th}$ harmonic | 40$^{th}$ harmonic | 60$^{th}$ harmonic |
|---|---|---|---|
| Seed laser wavelength | 262nm | 262nm | 262nm |
| Seed laser power | 1.2GW | 1.5GW | 1.5GW |
| Seed laser radius | 0.6mm | 0.6mm | 0.6mm |
| Modulator period | 0.1m | 0.1m | 0.1m |
| Modulator length | 0.4m | 0.4m | 0.4m |
| Modulator K | 4.2031 | 6.1753 | 7.5984 |
| R56$^{(1)}$ | 1.08mm | 2.86mm | 5.25mm |
| R56$^{(2)}$ | 0.0576mm | 0.0730mm | 0.0888mm |
| Beam energy | 0.7GeV | 1.0GeV | 1.22GeV |
| Radiator period | 29mm | 29mm | 29mm |
| Radiator K | 1.1793 | 1.1793 | 1.1793 |
| Radiation wavelength | 13.1nm | 6.55nm | 4.37nm |

### a) *20th harmonic case*

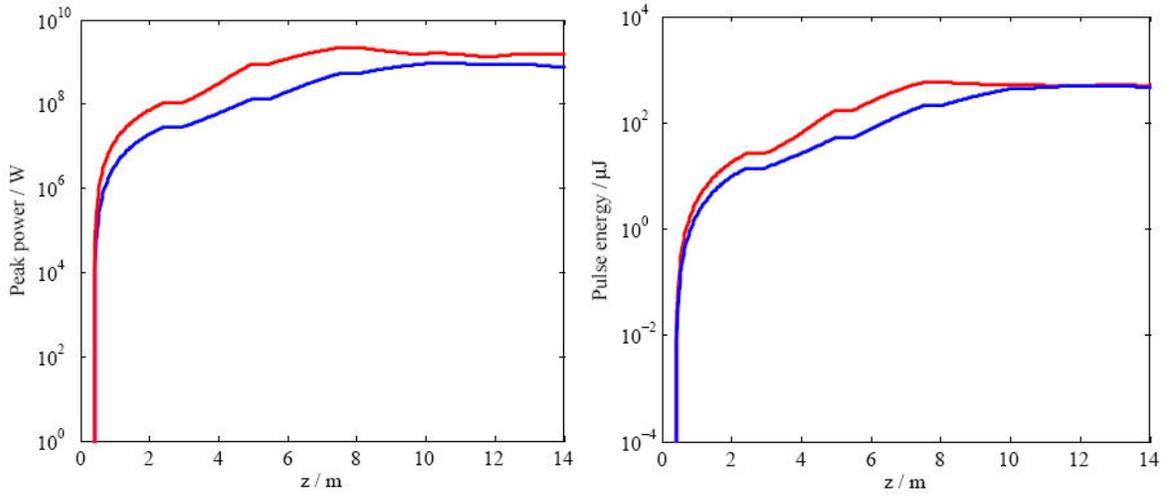

**Figure 28** The peak power and the pulse energy growth of 13.1nm radiation in the radiator, where the blue is the 1.25kA peak current case and the red is the 2.5kA peak current case.

In this section, we calculate the FEL performance of the EEHG-modulated beam in the radiator undulator of FLASH II. All the simulations include the CSR effects as described above. The same undulator as in FLASH is assumed. Table 3 lists the main parameters of EEHG option for FLASH II.

Figure 28 shows the peak power growth and the pulse energy growth along the radiator. The $20^{th}$ harmonic of the 262nm see laser, i.e. 13.1nm radiation, exceeds 0.7GW and 2GW in the 1.25kA peak current and 2.5kA peak current case, respectively. The pulse energy is about 200~300uJ in both cases. And Figure 29 shows the radiation power and the radiation phase distribution of the 13.1 nm pulse

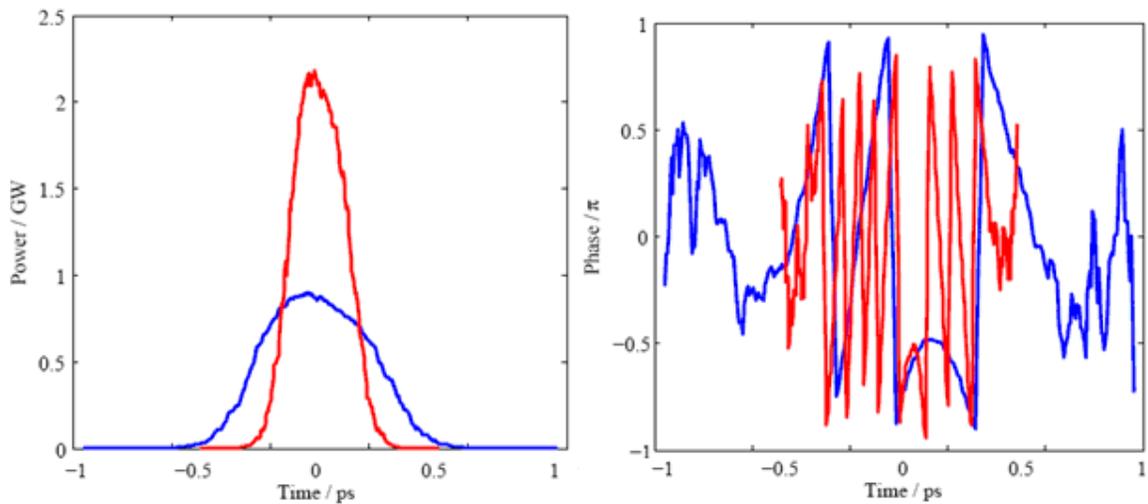

**Figure 29** The saturated radiation power pulse and radiation phase of the $20^{th}$ harmonic option for FLASH II, where the blue is the 1.25kA peak current case and the red is the 2.5kA peak current case.

### b) 40th harmonic case

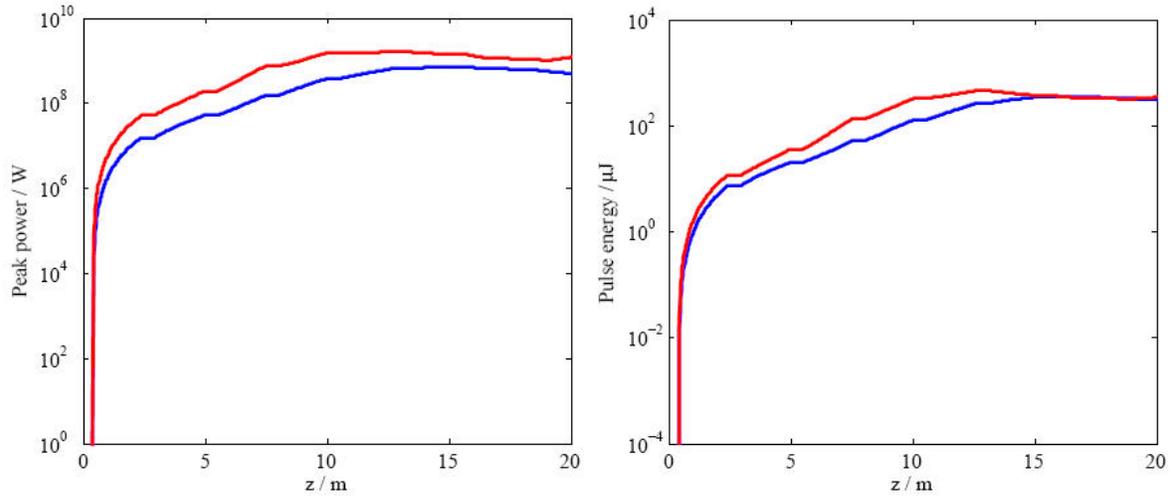

**Figure 30** The peak power and the pulse energy growth of 6.55nm radiation in the radiator, where the blue is the 1.25kA peak current case and the red is the 2.5kA peak current case.

Figure 30 shows the peak power growth and the pulse energy growth in the radiator. The 40th harmonic of the 262nm see laser, i.e. 6.55nm radiation exceeds 0.6GW and 1.6GW in the 1.25kA peak current and 2.5kA peak current case, respectively. The pulse energy is about 200uJ in both cases. And Figure 31 shows the radiation power and the radiation phase distribution of the 6.55 nm pulse.

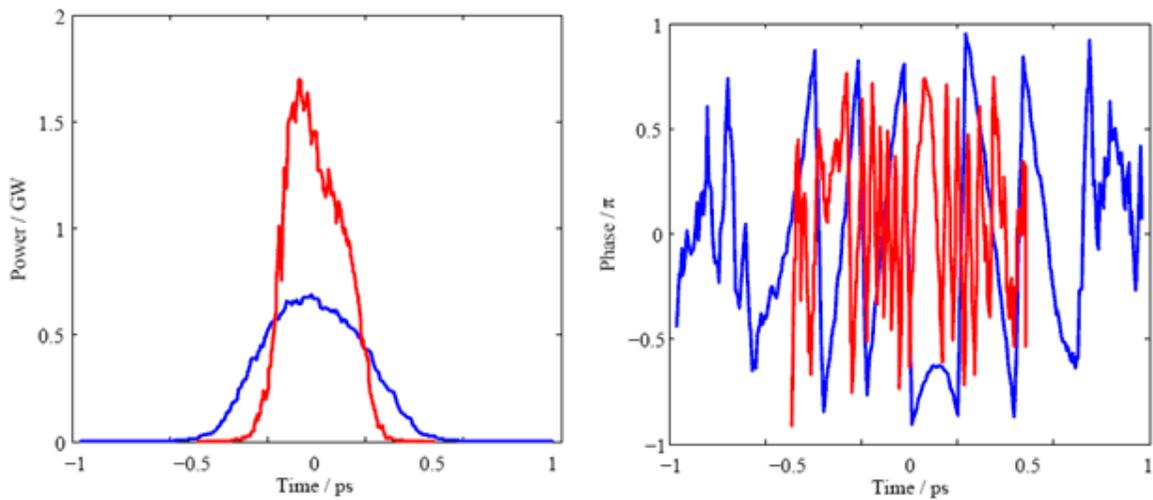

**Figure 31** The saturated radiation power pulse and radiation phase of the 40th harmonic option for FLASH II, where the blue is the 1.25kA peak current case and the red is the 2.5kA peak current case.

*c) 60th harmonic case*

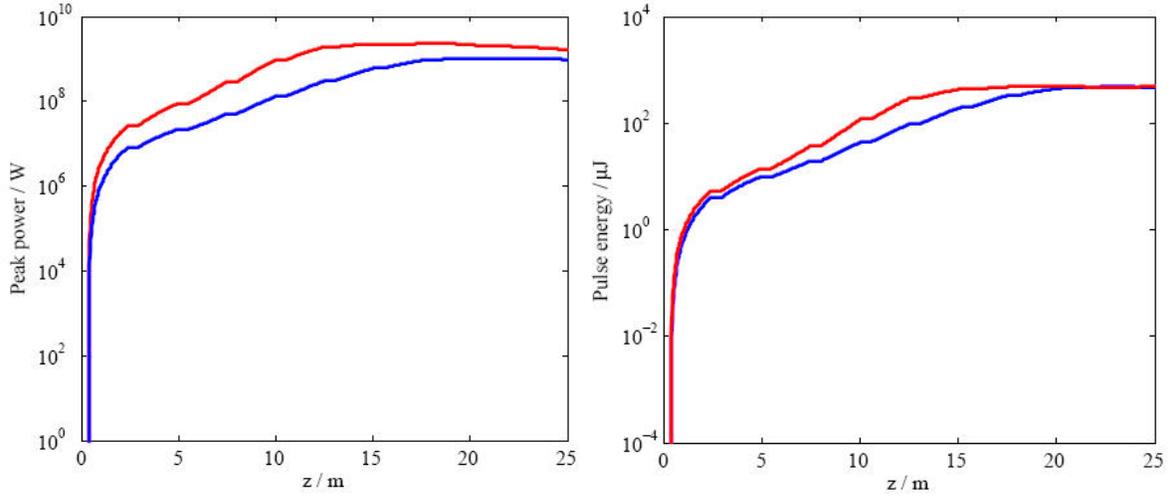

**Figure 32** The peak power and the pulse energy growth of 4.37nm radiation in the radiator, where the blue is the 1.25kA peak current case and the red is the 2.5kA peak current case.

Figure 32 shows the peak power growth and the pulse energy growth in the radiator. The 60th harmonic of the 262nm see laser, i.e. 4.37nm radiation exceeds 1GW and 2GW in the 1.25kA peak current and 2.5kA peak current case, respectively. The pulse energy is about 200~300uJ in both cases. And Figure 33 shows the radiation power and the radiation phase distribution of the 4.37 nm pulse.

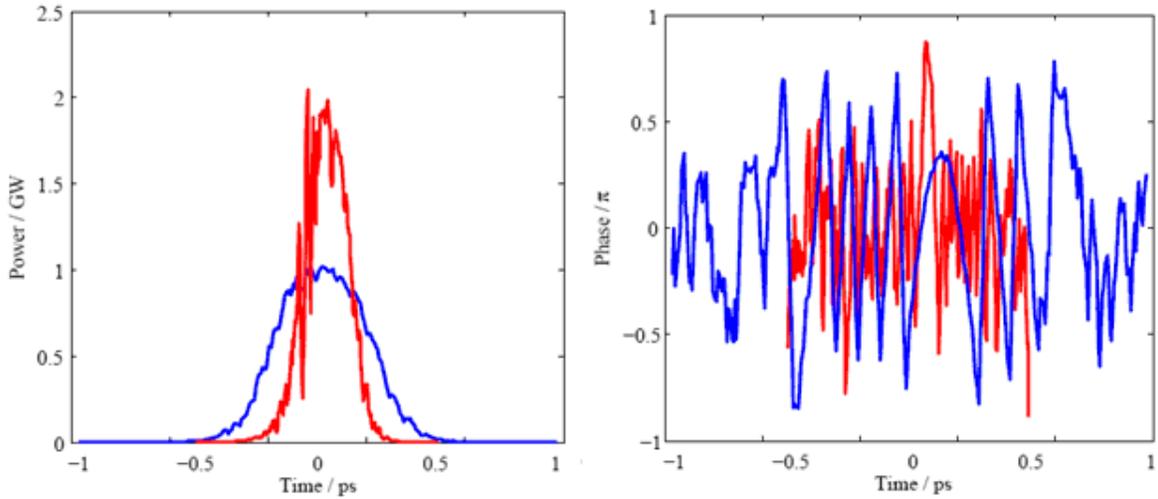

**Figure 33** The saturated radiation power pulse and radiation phase of the 60th harmonic option for FLASH II, where the blue is the 1.25kA peak current case and the red is the 2.5kA peak current case.

For FEL radiation, it is not the bunching factor of the whole electron beam, but the effective local bunching factor (i.e. the relevant sliced bunching factor) which determines the power growth in the radiator. As mentioned in section IV, the CSR broadens the microbunching bandwidth and reduces the microbunching amplitude of the whole beam. However, as seen in the simulation results, the sliced bunching factor does not degrade as much, and still sustains a significant coherent harmonic generation at the beginning of the radiator.

The sliced bunching factor correlates with the energy modulation induced by the CSR. It is seen from Figure 24 that the electron beam center experiences a larger energy variation, resulting in a lower bunching factor than other parts of the beam (see Figure 32). In our simulations we assume the seed laser to have an infinitely long flat-top longitudinal distribution. In practice the seed laser pulse length could be 10 ~ 100 fs and one could shift the seed laser pulse to select the beam parts with high density modulation efficiency.

Finally, we take a look at the radiation longitudinal coherence of the EEHG options for FLASH II. From the radiation pulse and radiation phase information given above, one can definitely state the existence of the longitudinal coherence, which will be further addressed in the next section.

## VI.  COMMENTS & DISCUSSIONS

In this section, several interesting issues for the EEHG simulations are discussed or commented. We take the 2.5kA peak current case, 13.1nm EEHG option for FLASH II as an example.

### a) *Dependence on macro-particle numbers*

In the above-mentioned simulations, an electron beam is represented by 2M macro-particles. As seen from the results, this is enough for the calculation of the projected bunching factor, and a clear signal-to-noise-ratio is observed over the natural statistics of the 2M macro-particles. However, if one considers the performance of the sliced beam, there are only thousands of macro-particles in each slice on a 262 nm scale. According to the numerical experience, thousands of macro-particles are enough for a SASE FEL simulation. However, when it comes to a seeded FEL, especially an EEHG, much more macro-particles are necessary.

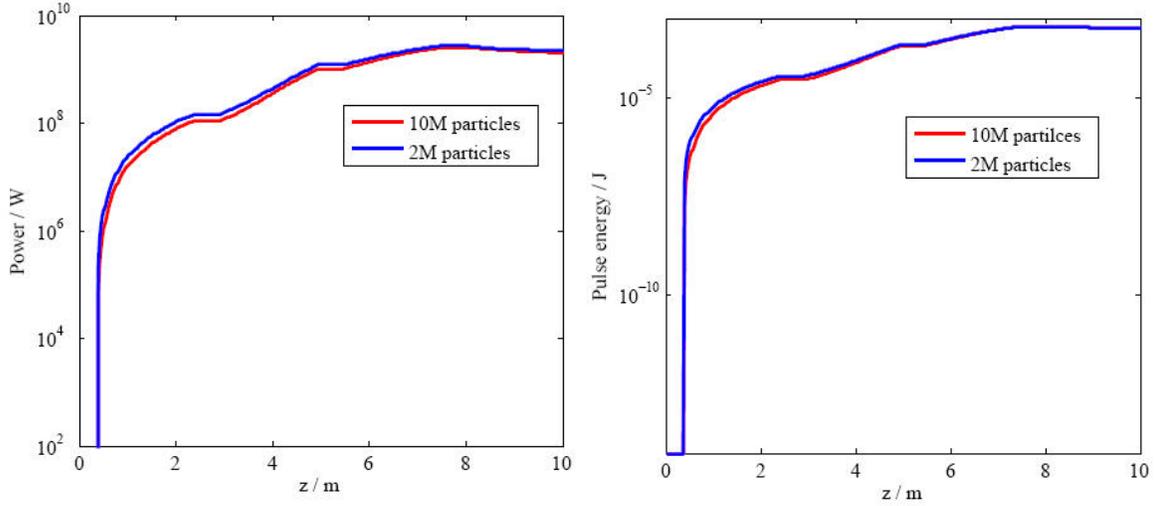

**Figure 34** The peak power and pulse energy growth of the 13.1 nm radiation in EEHG operation for FLASH II.

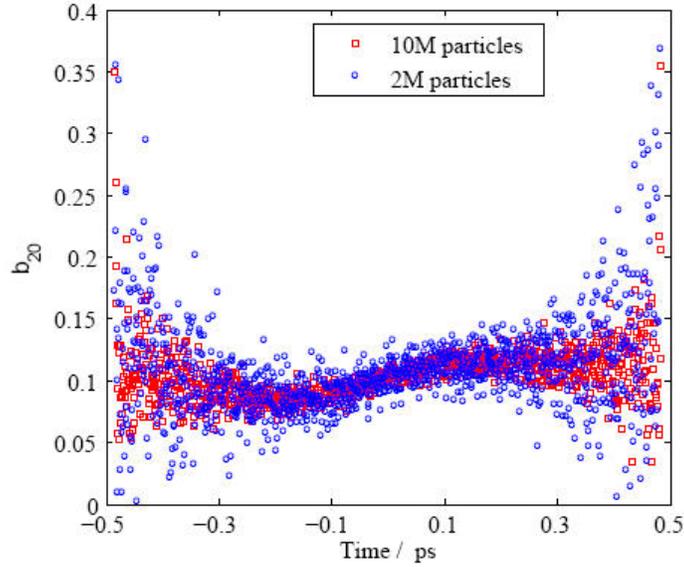

**Figure 35** The initial bunching factor of the 13.1 nm radiation in EEHG operation for FLASH II.

The maximum macro-particle number is limited by the required memory of the computer when the CSR effects are calculated. In our case, 10M macro-particles are almost the limit. In order to get an idea of the FEL parameters dependence on macro-particle numbers, the 10M macro-particles case is compared with the 2M macro-particles case.

Figures 34, 35 and 36 illustrated the corresponding results of 13.1nm EEHG radiation of FLASH II. The FEL performances of the 13.1nm EEHG radiation is almost similar in both cases, while the fluctuations of the initial bunching factor, saturated radiation power and saturated radiation

phase from slice to slice become smaller in the 10M macro-particles situation. It indicates that even more macro-particles will be helpful for the characterization of the longitudinal coherence and shot-to-shot fluctuation of EEHG radiation.

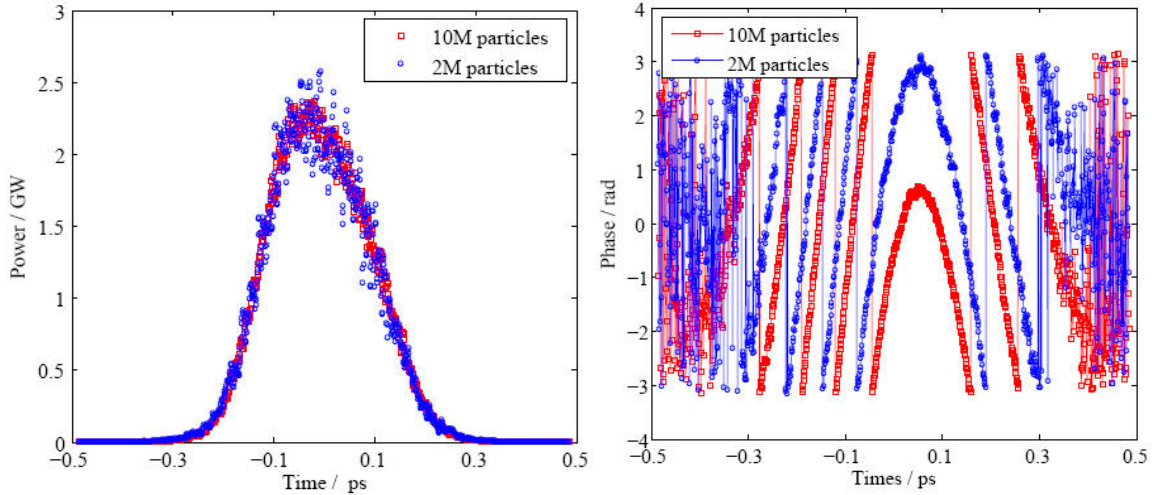

**Figure 36** The saturated radiation power and the radiation phase of the 13.1 nm radiation in EEHG operation for FLASH II.

### b) *Longitudinal coherence*

It is hard to accurately reproduce the radiation spectrum out of the simulations: the macro-particle numbers were limited to 10M by the memory of the used computer and the initial shot noise level [26, 27] for a longitudinal Gaussian beam can not be easily preserved in a numerical model (in this simulation the initial longitudinal distribution is just loaded from ASTRA generator [28]). Here we try to reproduce the radiation spectrum from the Fourier transform of the radiation pulse, and analyze the information obtained from the radiation spectra.

Figure 37 gives the transformed radiation spectrum where the CSR effects are not included. Since the EEHG fine structures are not destroyed, the longitudinal coherence of the 13.1nm radiation is only determined by the 262nm seed laser and preserved. Thus, the whole 13.1nm radiation pulse is close to Fourier-Transform-Limited. The quasi-*Sinc* spectrum of the central 100 radiation slices is from a quasi flat-top pulse shape in the time domain, i.e. the central 87fs of the pulse.

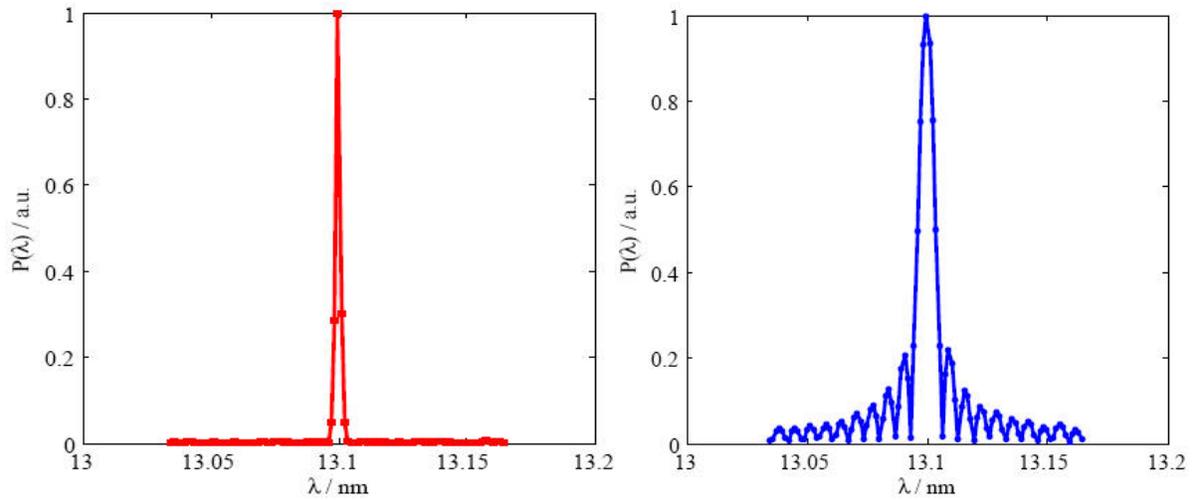

**Figure 37** The radiation spectrum of the 13.1 nm radiation in EEHG operation for FLASH II. The red is the spectrum from the whole radiation pulse and the blue is the spectrum from the central 100 slices of the radiation pulse. CSR effects are not taken into account in the results.

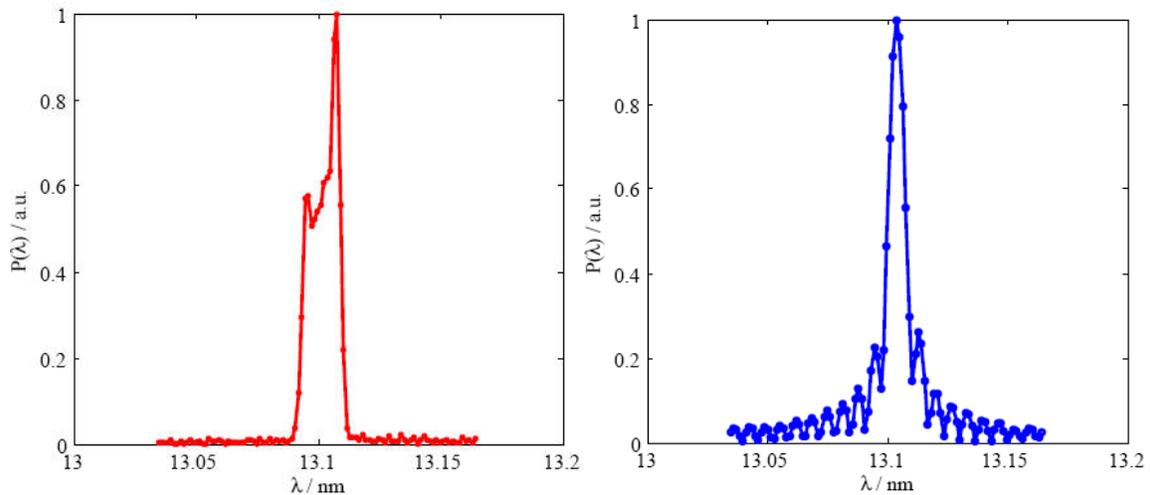

**Figure 38** The radiation spectrum of the 13.1 nm radiation in EEHG operation for FLASH II. The red is the spectrum from the whole radiation pulse and the blue is the spectrum from the central 100 slices of the radiation pulse. CSR effects are taken into account in the results.

Figure 38 gives the transformed radiation spectrum where the CSR effects are included. Since the EEHG fine structures are destroyed by the CSR field to some extent, the longitudinal coherence of the 13.1nm radiation can not be perfectly preserved. The bandwidth of the whole 13.1nm radiation pulse is broadened about 6 times. However, the central 100 slices radiation pulse is still close to Fourier-Transform-Limited. From this point of view, with a 10 ~ 100fs seed laser; there is always some steady working regime of the EEHG setup for FLASH II.

In order to compare the FEL performance between the EEHG option and a SASE case for FLASH

II, in Figure 39, the beam current distribution and the 13.1nm SASE spectrum are plotted. One can definitely conclude the existence of longitudinal coherence in EEHG option when compared with a SASE situation.

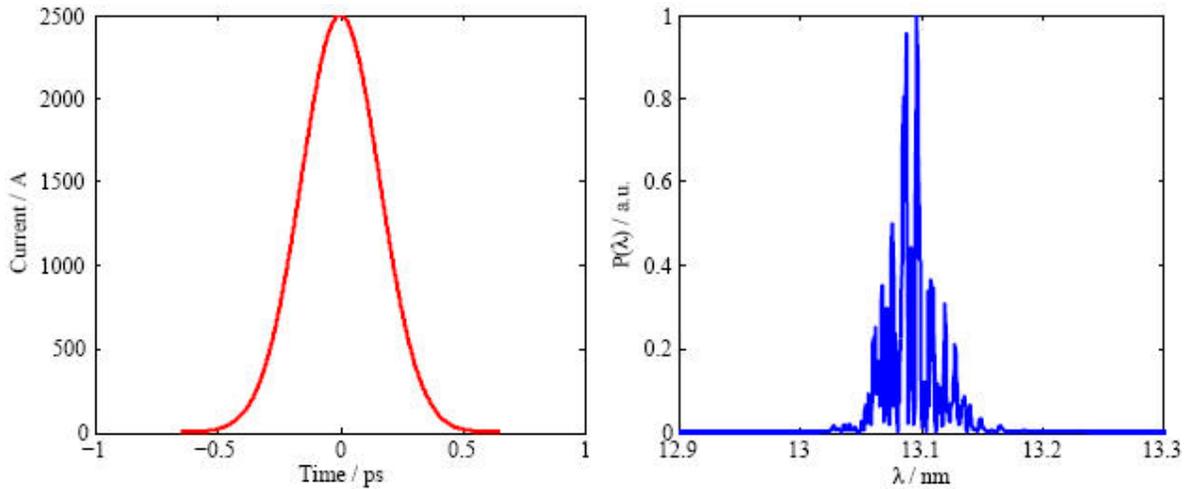

**Figure 39** The beam current distribution and the radiation spectrum of the 13.1 nm radiation in SASE operation for FLASH II.

c) *Dependence on the second chicane*

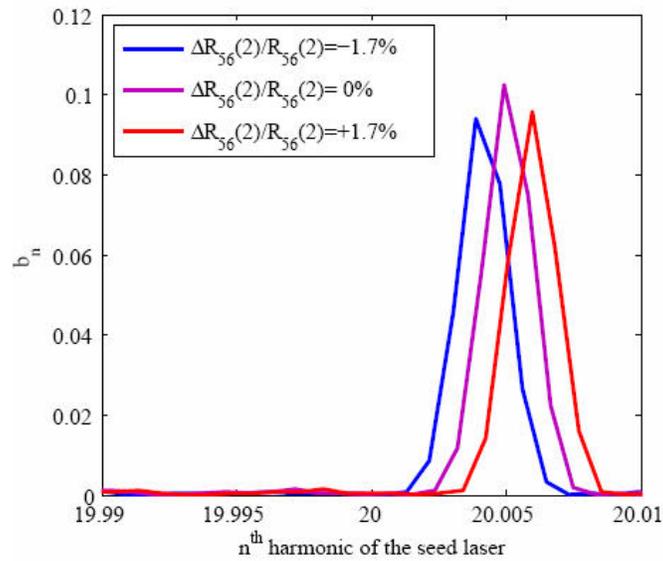

**Figure 40** The EEHG microbunching wavelength shift .vs. the dispersion of the $2^{nd}$ chicane when a negative linear energy chirp is introduced in the electron beam. CSR effects are not included.

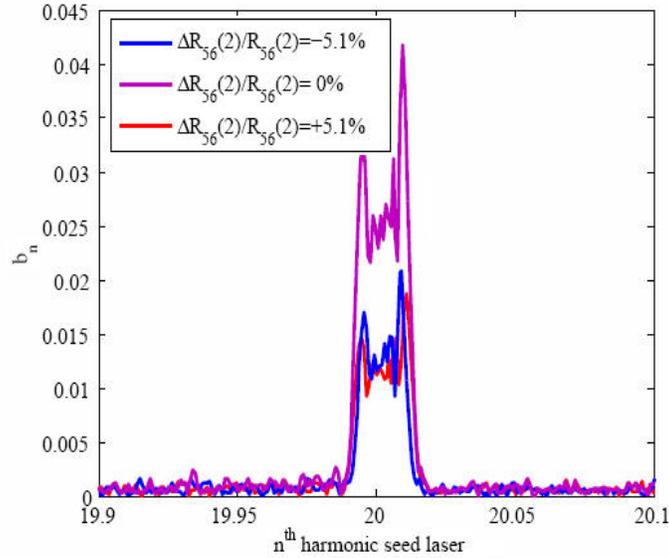

**Figure 41** The 13.1nm projected bunching factor .vs. the dispersion of the $2^{nd}$ chicane when the CSR effects in the $1^{st}$ chicane are taken into accounts.

The $2^{nd}$ chicane is small, but the EEHG performance is sensitive to the dispersion of the $2^{nd}$ chicane. Here some tolerance studies are presented. Figure 40 shows us the EEHG microbunching wavelength shift when the dispersion of the $2^{nd}$ chicane is varied in a ±1.7% range. Figure 41 shows the projected bunching factor when the dispersion of the $2^{nd}$ chicane is varied in a ±5.1% range.

### d) Benchmark with CSRtrack

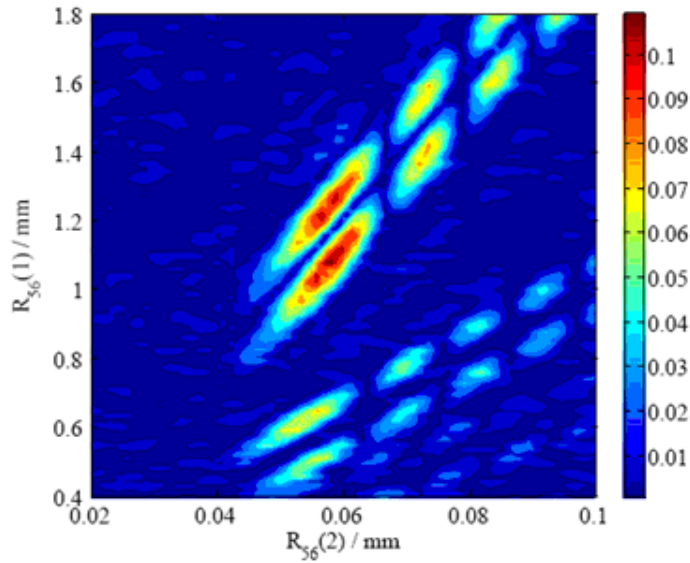

**Figure 42** The $20^{th}$ harmonic bunching .vs. the dispersion of the $1^{st}$ chicane and the $2^{nd}$ chicane in EEHG setup of FLASH II.

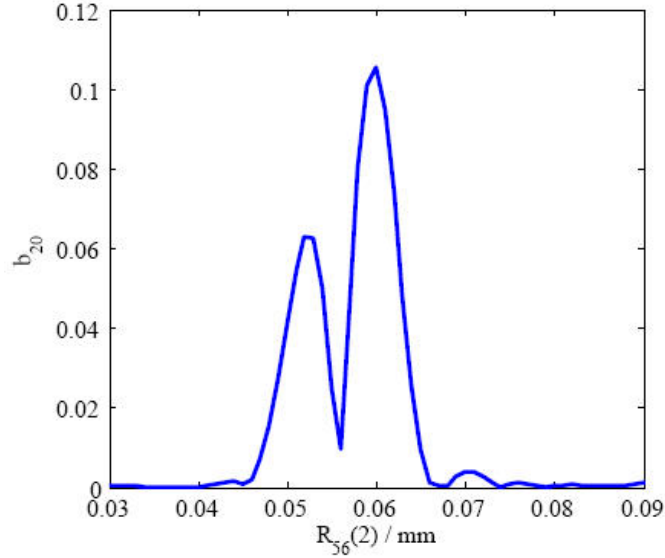

**Figure 43** The 20$^{th}$ harmonic bunching .vs. the dispersion of the 2$^{nd}$ chicane in EEHG setup, when the bending angle of the 1$^{st}$ chicane is optimized.

The beam dynamics & CSR effects in the 1$^{st}$ chicane was simulated by ELEGANT [29] in this study. As a benchmark, we use CSRtrack [30] to study the 13.1nm case. Figure 42 and 43 illustrate the 13.1nm bunching factor of the electron beam as the chicane parameters vary. Similar results as in ELEGANT are obtained by CSRtrack. The 13.1 nm microbunching is optimized at 1.12mm $R_{56}^{(1)}$ and 0.060mm $R_{56}^{(2)}$.

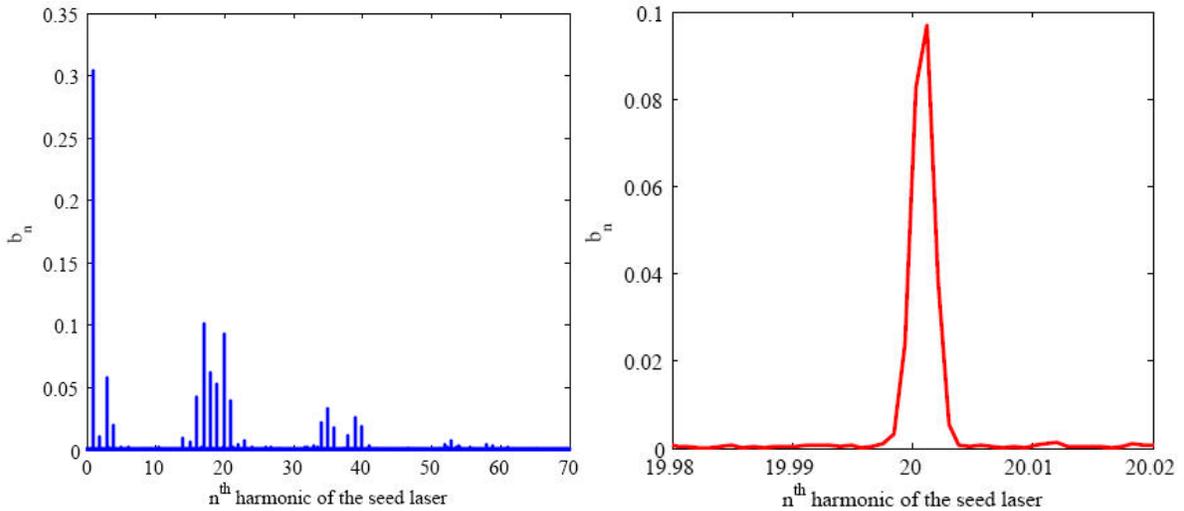

**Figure 44** The current spectrum of the modulated electron beam when EEHG setup is optimized at the 20$^{th}$ harmonic of the seed laser.

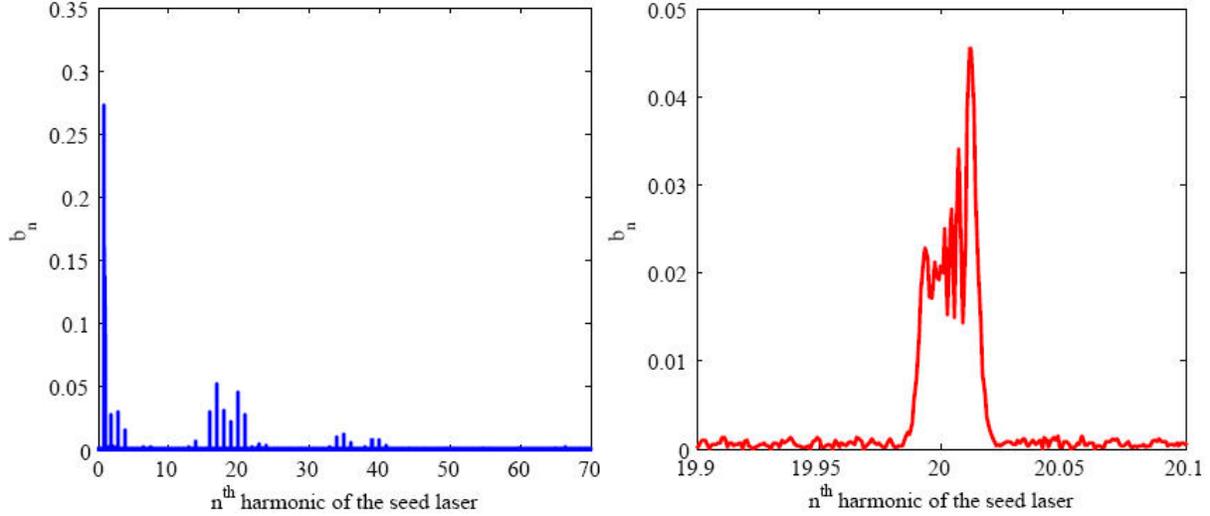

**Figure 45** EEHG microbunching spectra of the 20th harmonic EEHG option for FLASH II, where CSR effects in the 1st chicane are taken into account by CSRtrack.

Figure 44 and 45 gives the beam current spectrum, when the electron beam is modulated by the optimized EEHG setup. With the presence of the 'projected' type CSR force, the bandwidth of the projected bunching factor is broadened and the amplitude of the projected bunching factor degraded. It is in reasonable agreement with the ELEGANT results. Some differences are observed between the ELEGANT results and the CSRtrack results, which may be attributed to numerical errors.

*e) EEHG fine structures*

Figure 46 shows the fine structures of EEHG modulated electron beam. It is seen that CSR effects do not destroy the EEHG structures on the seed laser wavelength scale. However, whether the CSR algorithm of ELEGANT and CSRtrack are valid for this nm scale structures is a topic of additional future studies.

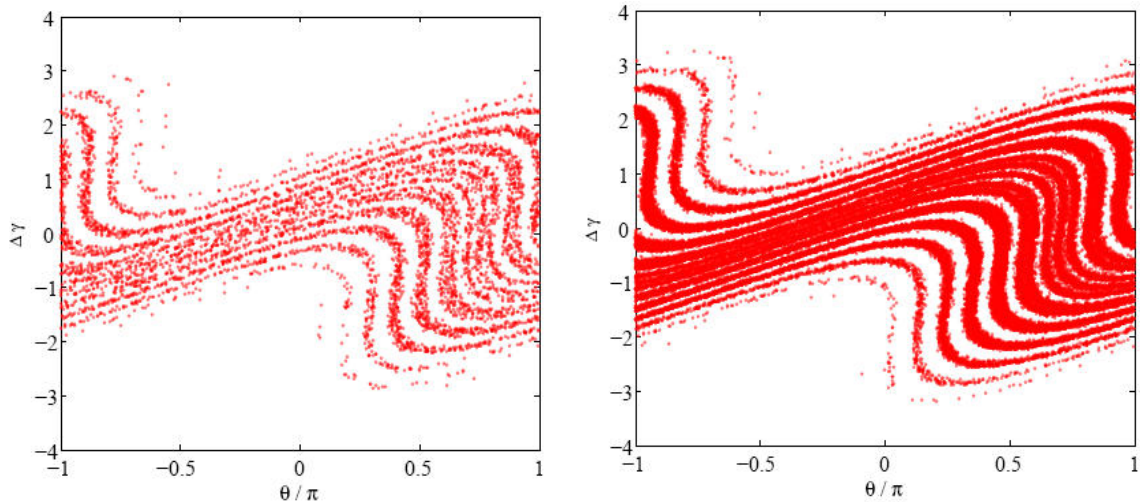

**Figure 46** The fine structure of EEHG modulated beam. The left and the right are from the beam

head and the beam center, respectively.

### *f) Microbunching instability*

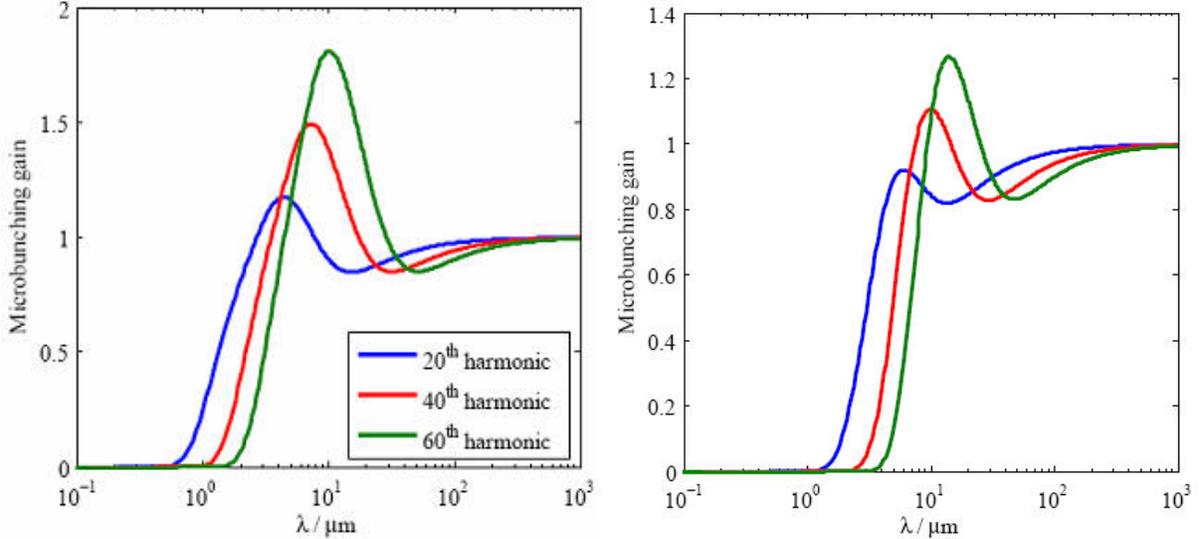

**Figure 47** The microbunching instability gain induced by CSR in the 1st chicane of EEHG option for FLASH II. The sliced energy spread in the left and the right is 0.2MeV and 0.45MeV, respectively.

The FLASH Linac delivers a low emittance (1.5mm-mrad), high peak current (2.5kA) electron beam to the undulator system. Thus, CSR-induced microbunching instabilities in the 1st chicane of the EEHG setup are of great interest. The mechanism of the CSR-induced microbunching instability amplification is well understood [31], and analytical solutions [31] are well validated by numerical simulations [32] and the LCLS laser-heater experience [33].

Here we analytically estimate the microbunching gain in the 1st chicane of the EEHG options for FLASH II. Figure 47 illustrates the gain of the microbunching from 0.1μm to 1000μm wavelength range. Since the absolute sliced energy spread of FLASH is 0.2MeV, and more energy modulation is induced in the 1st modulator of EEHG, the microbunching gain is well controlled to be fewer than 2 in a large wavelength range.

The microbunching instabilities study was motivated to see if the EEHG fine structures can be preserved in the presence of the CSR-induced microbunching instabilities. Considering a typical longitudinal phase space of the EEHG mechanism (see Figure 1), if one assumes that the electron energy is uniformly distributed in every fine structure, then on the seed laser wavelength scale, the RMS energy spread of each fine structure for FLASH II case can be analytically estimated as

$$\delta_{fine}[MeV] \approx E[GeV]/R_{56}[mm]/10 .$$

Then the RMS energy spread of the fine structure is 65keV, 35keV and 23keV in the 20$^{th}$, 40$^{th}$ and 60$^{th}$ harmonic EEHG setup for FLASH II, respectively. When we take the energy spread of the fine structure as the criteria in the reference [31], Figure 48 shows the microbunching instabilities gain from 0.1μm to 1000μm wavelength range. Since the fine structure was formed after the 1$^{st}$ chicane, thus the results in Figure 47 should be the worst-case estimate.

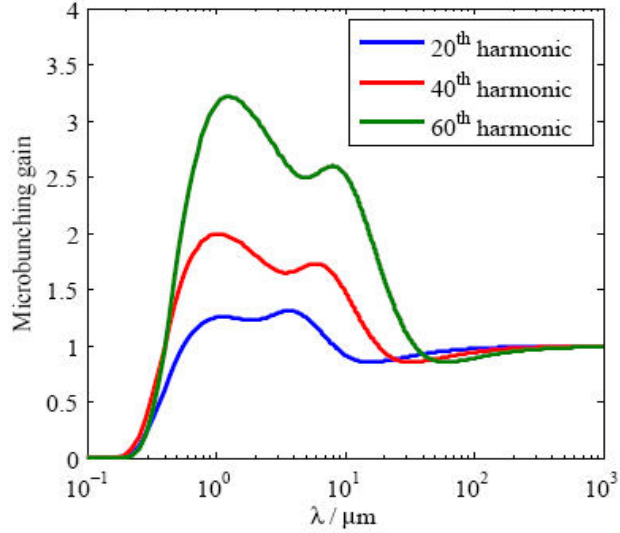

**Figure 48** The microbunching instability gain induced by CSR in the 1$^{st}$ chicane of EEHG option for FLASH II. The sliced energy spread of the fine structure is taken to be the criteria.

*g) Comparison with EEHG theory*

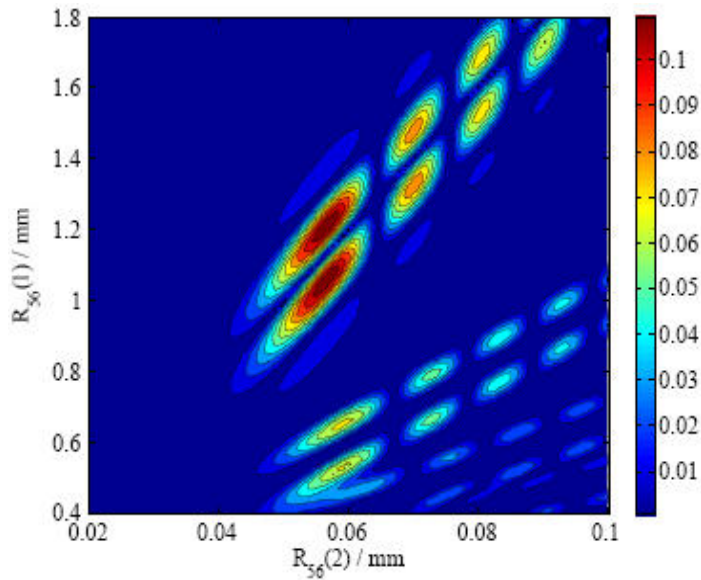

**Figure 49** The theoretical optimization of the 13.1nm EEHG option for FLASH II, in which the normalized energy modulation parameter is chosen as 3 in modulator 1 and 2.

EEHG theory predicts an analytical solution for the bunching factor. The analytical optimization of 13.1nm EEHG option for FLAH II is shown in Figure 49, from which a good agreement between the EEHG theory and the numerical simulation is observed.

### *h) Issues on the seed laser noises*

The noise propagation issue in harmonic generation FEL is of great interest, and was originally stressed for an HGHG FEL [34~35]. For an EEHG FEL, there is some theoretical investigation on the noise issue [36~37]. Here, we present some brief numerical results of the noise issue in EEHG option for FLASH II. The 13.1nm EEHG radiation at the saturation point in the time domain and in the spectral domain is shown in Figure 50, where a seed laser noise with 1% RMS power error and 1 degree RMS phase error is introduced. Additional output jitter of the 13.1nm radiation from the 262nm seed laser noise is clearly observed.

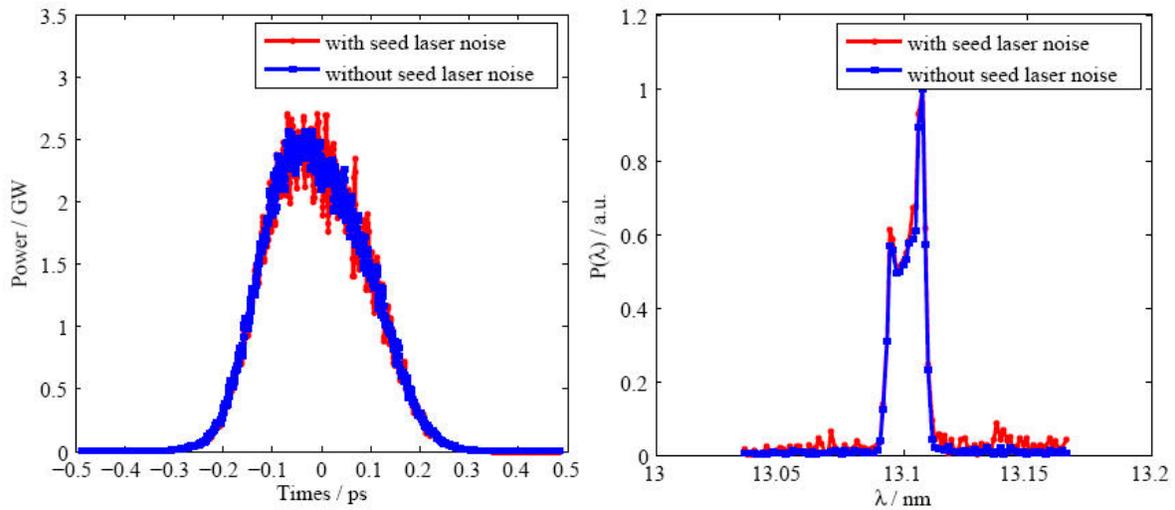

**Figure 50** The radiation pulse and radiation spectrum of the 13.1nm EEHG option for FLASH II, in which the CSR effects are included.

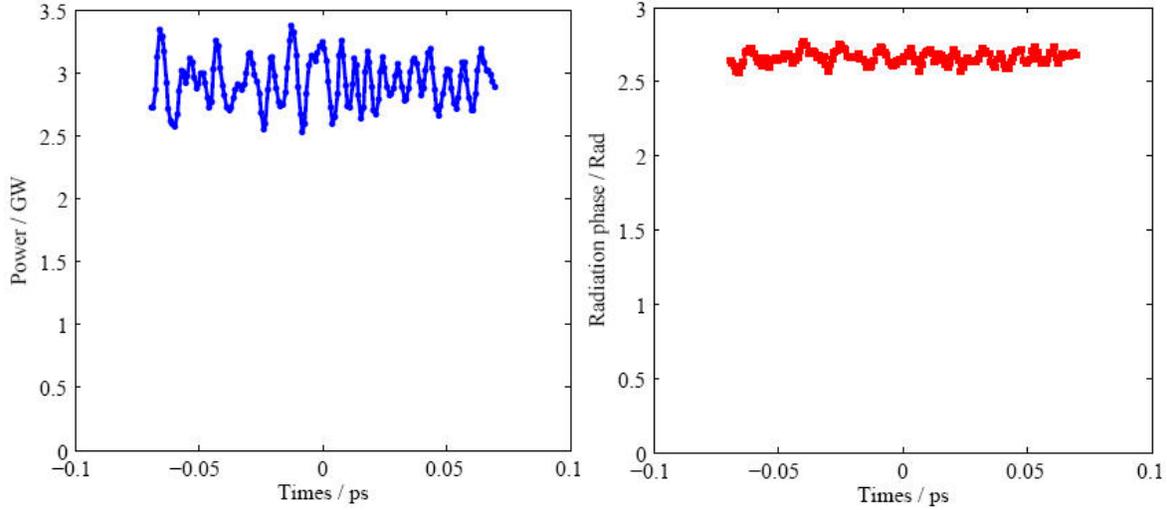

**Figure 51** Jitter of the radiation power and radiation phase of the 13.1nm EEHG option for FLASH II, where only the shot noise effects of the electron beam are taken into accounts.

The chaotic output in Figure 50 is attributed to the CSR effect, the electron beam shot noise, the seed laser noise (both amplitude and phase) and the numerical errors. In order to effectively remove the unknown numerical errors in a Gaussian beam description, a flat-top infinite long electron beam is utilized, where the shot noise level of the electron beam are preserved [27]. Moreover, the CSR effects are not included in the following steps.

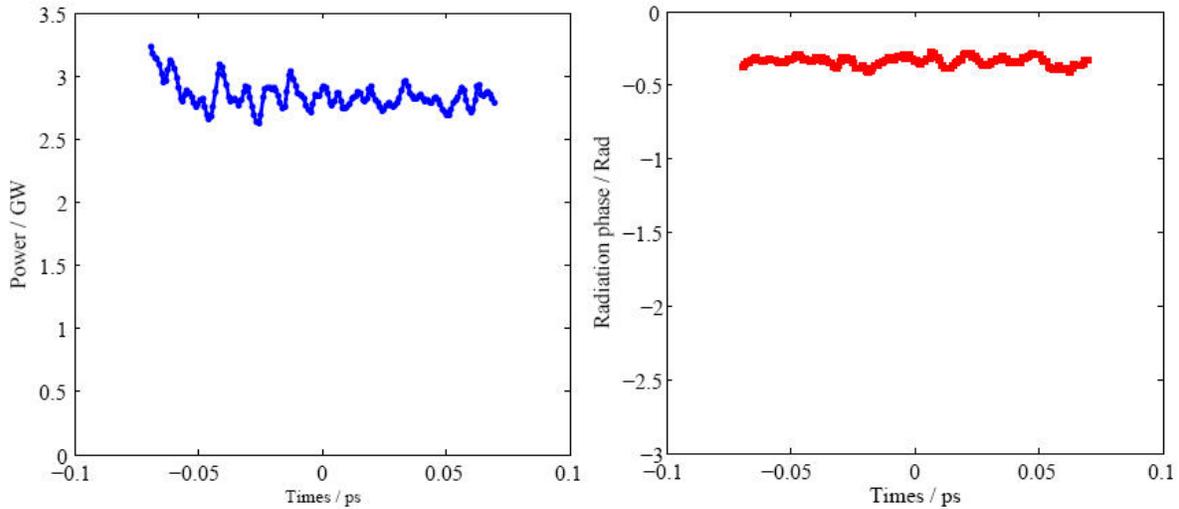

**Figure 52** Jitter of the radiation power and radiation phase of the 13.1nm EEHG option for FLASH II, where only a 1% RMS error of the seed laser power is induced.

Figure 51, 52 and 53 show the effects of the electron beam shot noise, the seed laser amplitude noise (1% RMS) and the phase noise (1 degree RMS) on the 13.1nm output radiation power and radiation phase, respectively.

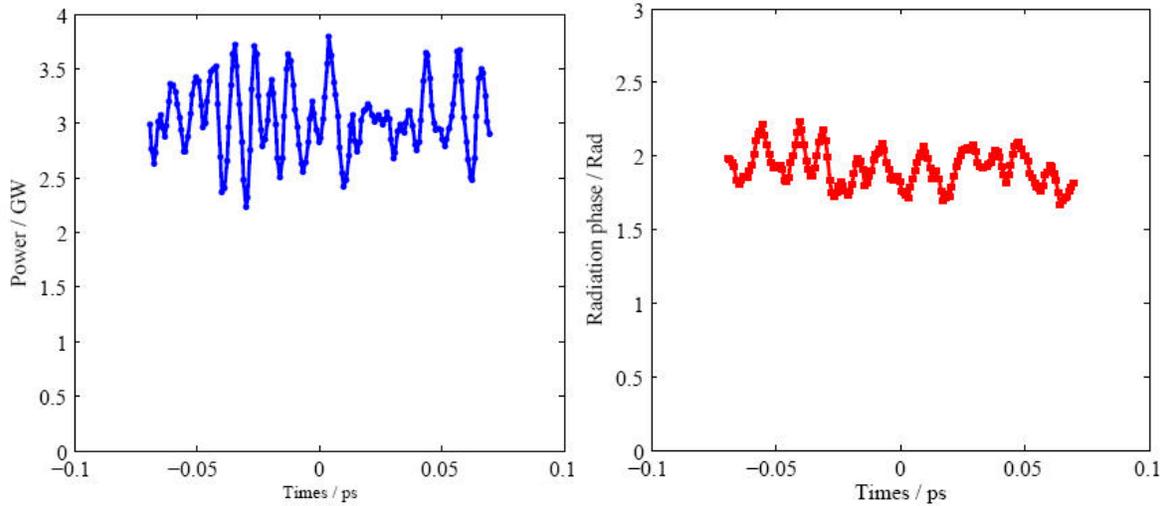

**Figure 53** Jitter of the radiation power and radiation phase of the 13.1nm EEHG option for FLASH II, where only a 1 degree RMS error of the seed laser phase is induced.

Figure 54 shows the 13.1nm FEL output performance at different magnitudes of the seed laser noise. It is seen that the FEL output performance degrades monotonously with the increase of the seed laser noise, and the seed laser phase error contributes more than the seed laser amplitude noise.

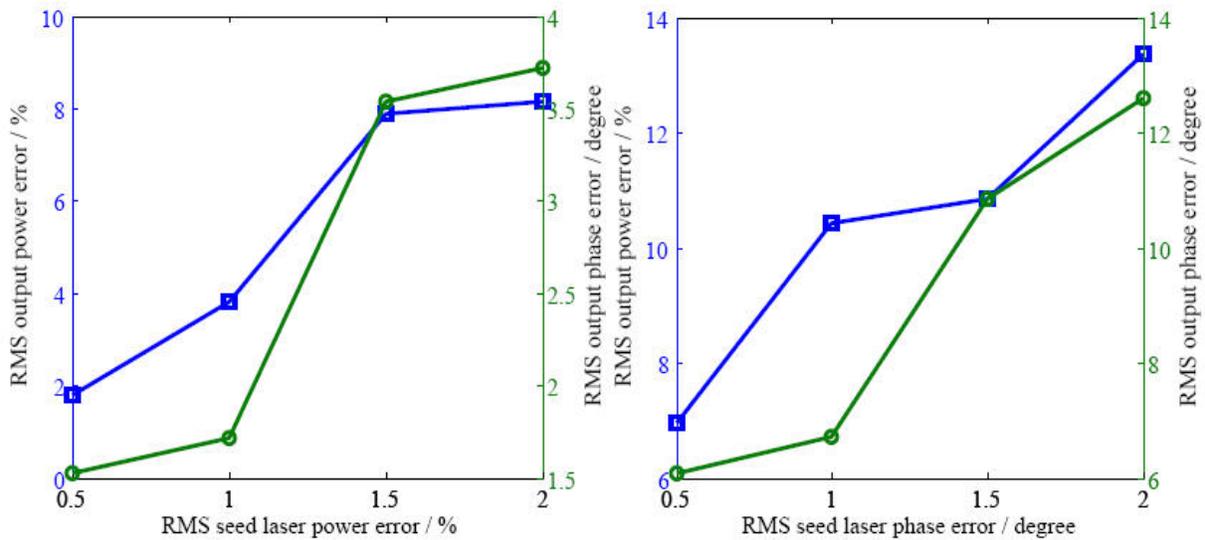

**Figure 54** The FEL performance of the 13.1nm EEHG option for FLASH II .vs. the 262 nm seed laser noise.

Figure 55 presents the 13.1nm FEL output performance with the electron beam shot noise and the seed laser noise included simultaneously.

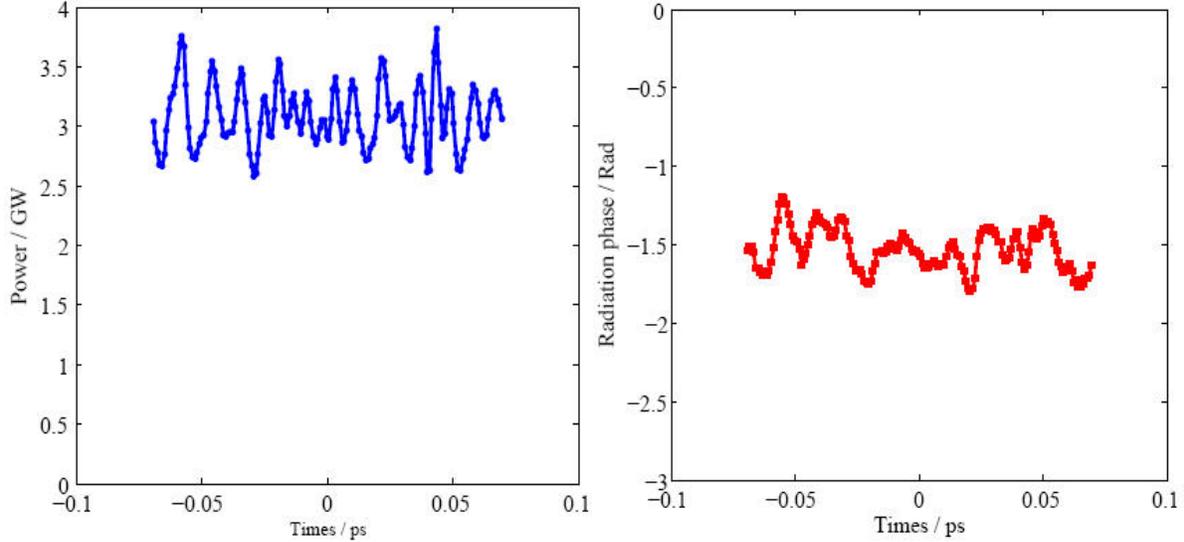

**Figure 55** Jitter of the radiation power and radiation phase of the 13.1nm EEHG option for FLASH II, where the seed laser noise and the electron beam shot noise are induced.

## VII.    CONCLUSIONS

The echo-enabled harmonic generation scheme is considered to be a potential operation mode for FLASH II which is an upgraded seeded free electron laser project of FLASH. In this paper, the EEHG option is numerically investigated and optimized for FLASH II. The effects of the beam energy chirp and the coherent synchrotron radiation are also extensively simulated. It is shown that EEHG is very robust at several tens harmonics of the seed laser. Moreover, several discussions corresponding to the EEHG simulations are presented. It is worth stressing that this work is preliminary and there is still room for further improvement. Benchmarks from theories, other codes and experiments, extensive tolerance studies and start2end simulation should be the future work.

## ACKNOWLEDGMENT

We are grateful to Jianhui Chen, Meng Zhang, Bo Liu, Qiang Gu, Dong Wang and Zhimin Dai from SINAP, Matthias Scholz, Torsten Limberg and Martin Dohlus from DESY, Sven Reiche from PSI, Yuantao Ding, Juhao Wu and Zhirong Huang from SLAC, Yuhui Li from European XFEL and Jun Yan from Duke University for helpful discussions.

## APPENDIX A: Laser Beam Interaction Code in Undulator (LBICU)

Suppose a planar undulator with a sinusoidal magnetic field in the *y* direction and a period length $\lambda_u$ in the *z* direction. Considering a relativistic electron beam with average energy $\gamma_0 mc^2$ and a coherent laser with wavelength $\lambda_s$ enter the planar undulator together, one may observe the electron's transverse wiggling motion and the longitudinal "figure-eight" oscillation. Such a trajectory gives rise to energy exchange between the electron beam and the laser electric field. We denote the wave numbers of the seed laser and the undulator magnets by $k_s = 2\pi/\lambda_s$ and $k_u = 2\pi/\lambda_u$, respectively. Then, the magnetic field distribution of the planar undulator can be written as

$$B_y = B_0 \sin k_u z, \qquad (1)$$

where $B_0$ is the undulator peak magnetic field. And the electric field of a seed laser with Gaussian distribution and the *rms* size of $\sigma_x$, $\sigma_y$ and $\sigma_z$ can be represented as

$$E_x^2 = E_0^2 \sin^2[k_s(z'-z_0)+\varphi_0] e^{-\frac{x^2}{2\sigma_x^2}-\frac{y^2}{2\sigma_y^2}-\frac{(z'-z_0)^2}{2\sigma_z^2}}, \qquad (2)$$

where $z_0$ is the initial relative position of the laser from the electron beam, and $\varphi_0$ is the carrier envelope phase of the laser. The diffraction effects of the laser field can be approximated by

$$\sigma_x(z) = \sqrt{\sigma_{xw}^2 + \frac{k_s^2(z-z_w)^2}{4\sigma_{xw}^2}},$$
$$\sigma_y(z) = \sqrt{\sigma_{yw}^2 + \frac{k_s^2(z-z_w)^2}{4\sigma_{yw}^2}}, \qquad (3)$$

where $\sigma_{xw}$ and $\sigma_{yw}$ denote the laser size at the longitudinal waist position $z_w$. Then the electron beam's motion satisfies the law of electrodynamics,

$$\gamma m \frac{dv}{dt} = eE - ev \times B. \qquad (4)$$

In the presence of the transverse wiggling motion of the electron beam and the transverse field of the seed laser, an energy exchange between the electrons and the electromagnetic field is expected as

$$mc^2 \frac{d\gamma}{dt} = eE_x \frac{dx}{dt}. \qquad (5)$$

On the basis of Equations (1) ~ (5), a three-dimensional (3D), time-dependant code LBICU is developed to numerically study the laser-beam interaction in an undulator. The initial electron beam distribution is loaded from GENESIS [38] to remove the effect of finite electron numbers, and the integral step of each undulator period is set to be 100 for an accurate solution. To check the validity of the algorithm, for the case of infinitely long seed laser and infinitely long electron beam, several comparison studies have been carried out by using the steady-state mode of GENESIS and good agreement is observed.

Under such an approach, the detuning effects of an undulator, the carrier envelope phase effects of a few-cycle laser, wavelength detuning of seeded FEL [39] and even the laser-beam interaction in a dipole [40] can be easily modeled. It is worth stressing here that the seed laser evolution in short modulator is neglected in LBICU. Usually, the modulator can easily drive the FEL process at longer wavelength and thus slightly altering the energy modulation level after the modulator. Fortunately, EEHG mechanism prefers an extremely short modulator, and it is just 4 periods in the studied case.

### APPENDIX B: Beam Dynamics in Chicanes

The beam dynamics in the chicane, including ISR and CSR are tracked by the well benchmarked code ELEGANT. Moreover, other codes, e.g. CSRtrack and IMPACT [41], can easily substitute for ELEGANT in our numerical model.

### APPENDIX C: Self-Consistent FEL Simulation

The FEL simulation is carried out by GENESIS, with a self-consistent method proposed in ref. [42], where the full electron information from the beam dynamic tracking code is imported into the FEL code GENESIS.